# Fast Isotropic Li-Ion Diffusion in Zeolitic Imidazolate Framework Glass Electrolytes for Batteries

Yong Li [a], Tao Du [b,*], Timothée Jamin [a], Zhencai Li [a], Kasper Tolborg [a], Yuanzheng Yue [a], Morten M. Smedskjaer [a,*]

[a] *Department of Chemistry and Bioscience, Aalborg University, 9220 Aalborg East, Denmark*

[b] *Department of Applied Physics, The Hong Kong Polytechnic University, Kowloon, Hong Kong 999077, China*

[*] *Corresponding authors. E-mail: mos@bio.aau.dk (M.M.S.); dutaohit@gmail.com (T.D.)*

**Abstract:** All-solid-state lithium batteries require solid electrolytes that combine rapid room-temperature ion transport with mechanical robustness and interfacial compatibility. Zeolitic imidazolate framework (ZIF) glasses, with ZIFs being a sub-set of metal-organic frameworks, offer an attractive yet relatively underexplored platform because they combine an grain-boundary-free and amorphous topology with chemically tunable frameworks. Here, we reveal that structural disorder unlocks fast and isotropic lithium diffusion in ZIF glasses. This is realized by using a machine learning interatomic potential to simulate $Li^+$ transport in crystalline and glassy ZIF-4 and ZIF-62. Structural disorder reduces the activation energy for $Li^+$ migration from ~0.35 eV to 0.16 eV and increases the extrapolated room-temperature diffusion coefficient by more than one order of magnitude for ZIF-4 and nearly sevenfold for ZIF-62. Analyses of non-Gaussian dynamics and van Hove correlation functions reveal that $Li^+$ diffusion in crystalline ZIFs occurs via rare, dynamically heterogeneous hopping events among well-defined cages, whereas $Li^+$ diffusion in glassy ZIFs is more homogeneous, continuous, and Fickian-like, benefiting from a wide distribution of coordination geometries and migration barriers. $Li^+$ diffusion in crystalline ZIFs is strongly anisotropic, reflecting that ordered orientations of imidazolate and benzimidazolate rings impose distinct energy barriers along different crystallographic directions. Upon vitrification, these ring orientations become randomized, and hence, the diffusion of $Li^+$ becomes isotropic or near-isotropic. These findings imply that well-designed metal-organic framework glasses are a promising candidate as high-performance solid-state electrolytes.

## 1. INTRODUCTION

The practical energy densities of lithium-ion batteries, a cornerstone of modern energy storage, are approaching their limits, particularly for conventional graphite anode-liquid electrolyte systems.[1] At the same time, liquid electrolytes suffer from safety concerns, including flammability and leakage.[2] Importantly, the emerging all-solid-state lithium batteries offer a promising route toward high energy density and intrinsic safety. A central challenge, however, is the issues with the use of lithium metal anodes. Although solid electrolytes are often expected to suppress dendrite growth, lithium penetration and possible fracture can still occur even in solid electrolytes with high elastic moduli.[3, 4] This has shifted attention toward solid electrolytes that combine high ionic conductivity with mechanical resilience, interfacial compatibility, and resistance to failure. Recently, Kalnaus et al. emphasized that many critical failure modes in all-solid-state batteries are fundamentally mechanical in origin, highlighting the need to design the electrolyte that can regulate stress and strain evolution during cycling.[5]

Each of the current solid electrolyte families face their own intrinsic trade-offs. Sulfide electrolytes exhibit some of the highest room-temperature ionic conductivities, but their air sensitivity and limited interfacial chemical stability complicate practical implementation.[6] Oxide electrolytes offer superior chemical stability and wide electrochemical windows, yet typically suffer from poor electrode/electrolyte contact and high grain-boundary resistance to ionic conduction.[7] Polymer and polymer-composite electrolytes provide attractive processability and interfacial compliance, but their room-temperature ion transport is often limited by sluggish segmental dynamics, while their mechanical strength remains insufficient for robust lithium-metal operation.[8] Crystalline inorganic electrolytes provide well-defined ion-transport pathways but are often brittle and susceptible to grain-boundary-related failure, whereas polymeric systems offer mechanical compliance at the expense of ion-transport efficiency. No single class of solid electrolyte has yet fully satisfied the combined requirements of high ionic conductivity, fracture toughness, electrochemical stability, and stable interfaces.

Metal-organic framework (MOF) glasses, and particularly zeolitic imidazolate framework (ZIF) glasses, have recently emerged as a distinct subclass of melt-quenched glasses that could help bridge this gap. [9, 10] ZIF glasses are typically obtained by melt-quenching ZIF crystals, such as ZIF-4 and ZIF-62, thereby transforming long-range ordered frameworks into disordered ones while retaining a large degree of metal-ligand connectivity and framework porosity.[11] Their single-phase, grain-boundary-free structure distinguishes them from polycrystalline ceramics, while their rigid yet topologically disordered networks avoid the segmental-motion limitations inherent to polymer electrolytes. Beyond their established promise in gas separation,[12] optics,[13, 14] and energy applications,[15] ZIF glasses also exhibit unusual mechanical characteristics. ZIF-62, for example, exhibits exceptional glass-forming ability ($10^5$ Pa·s viscosity at the melting point),[16] low liquid fragility ($m$ = 23), relatively high Poisson's ratio,[17] and localized shear deformation under stress (e.g., shear bands).[18] Recently experiments have shown that ZIF glass layers can promote smooth lithium deposition, suppress dendrite growth, and improve cycling stability in lithium-metal batteries.[19] These features make ZIF glasses a promising platform for solid electrolytes, enabling the optimization of ion transport and mechanical stress accommodation through tailoring framework topology.

Despite this promise, the microscopic mechanisms of $Li^+$ transport in ZIF glasses remain poorly understood. It is unclear how the transition from structural order to disorder modifies the lithium migration landscape, activation barriers, transport heterogeneity, and directional anisotropy. Addressing these questions requires atomistic simulations to accurately capture both the flexible, chemically diverse framework and lithium-framework interactions. Classical force fields, such as the Universal Force Field, have been used for ZIF systems and offer computational efficiency, but they are limited in their ability to quantitatively describe local coordination dynamics, disordered glassy environments, and $Li^+$ migration.[20] Deep-learning interatomic potentials provide a promising alternative by ensuring density-functional-theory accuracy at near-classical molecular dynamics (MD) cost.[21, 22] Recently, DeePMD-based force fields have successfully reproduced structural transitions and mechanical behavior in crystalline and glassy ZIFs, including ZIF-4 and related systems.[23,24] Universal machine-learning potentials have also shown promise for MOF simulations,[25, 26] although their transferability to amorphous MOF structures remains an active area of development.

In the present work, we fine-tune a pre-existing DeePMD-based deep-learning potential[23], which we previously trained for ZIF glass systems, by incorporating Li-specific interaction data. We then use this refined potential to investigate $Li^+$ diffusion in crystalline and glassy ZIF-4 and ZIF-62 (see Methods section for details). These two representative frameworks differ in linker chemistry and in their glass-forming ability. By combining temperature-dependent diffusion calculations, activation-energy analysis, non-Gaussian dynamics, van Hove correlation functions, diffusion tensors, and linker-orientation descriptors, we establish how vitrification alters lithium transport at the atomic scale. We show that structural disorder lowers the $Li^+$ migration barrier, promotes more homogeneous room-temperature diffusion, and transforms strongly anisotropic crystalline transport into isotropic or near-isotropic diffusion in the glass. These findings reveal how framework disorder and linker-ring orientation govern $Li^+$ mobility in ZIF glasses and provide feasible principles for designing glassy MOF-based solid electrolytes.

## 2. RESULTS AND DISCUSSION

**2.1 Structural validation.** The structures of ZIF-4 and ZIF-62 consist of Zn-centered tetrahedra linked by imidazolate (Im), and benzimidazolate (bIm) and Im mixed ligands, respectively. Glassy phases were generated via melt-quenching (see Methods section), yielding densities of 1.57 g/cm$^3$ for ZIF-4 and 1.51 g/cm$^3$ for ZIF-62, in good agreement with experimental values (1.62 and 1.57 g/cm$^3$, respectively) and previous simulations (1.55 and 1.51 g/cm$^3$, respectively).[23, 27, 28] The lower density of ZIF-62 glass reflects a large free volume, due to the incorporation of bulkier bIm linkers into framework. To validate the structural models, pair distribution functions (PDFs) and structure factors were evaluated (Supporting Figure S1). The absence of significant peak shifts between crystalline and glassy phases confirms that Zn-N coordination and linker integrity are preserved, while vitrification primarily induces angular disorder of Im/bIm rings.[29, 30] The disappearance of sharp diffraction features and the emergence of broad peaks further indicate the loss of long-range order while maintaining short-range connectivity.

**2.2 Disorder-induced reorientation of linker rings.** The crystalline-to-glassy transition in ZIF-4 and ZIF-62 is driven by the accumulation of local angular distortions of Im/bIm rings, and the subsequently medium- to long-range rearrangements of Zn-Zn pairs. To probe this mechanism, we used the orientation propensity descriptor.[23] Originally developed to quantify Im-ring reorientation during the crystalline-to-glassy transition in ZIF-4, this descriptor is here extended to include bIm rings, enabling its application to ZIF-62. Figure 1*a* illustrates the workflow used to construct pole figures that capture the orientational preferences of Im and bIm rings on the *y*-*z*, *x*-*z*, and *x*-*y* planes. Ring normal vectors are extracted, projected onto a unit sphere, and then mapped onto three orthogonal planes by stereographic projection. Figure 1*b* shows the resulting pole figures for crystalline and glassy ZIF-4 and ZIF-62 solid electrolytes.

For crystalline ZIF-4, the pole figures display distinct projection-dependent patterns of density distribution in Im: a circular distribution with a hollow, curved-square central region on the *y*-*z* plane, a dog-bone-shaped central region on the *x*-*z* plane, and an elliptical central region on the *x*-*y* plane. These characteristic features reflect the anisotropic orientational ordering of Im rings within the crystalline lattice, where the rings preferentially align along specific crystallographic directions. The distinct shapes on different projection planes further highlight the three-dimensional nature of this orientational texture.

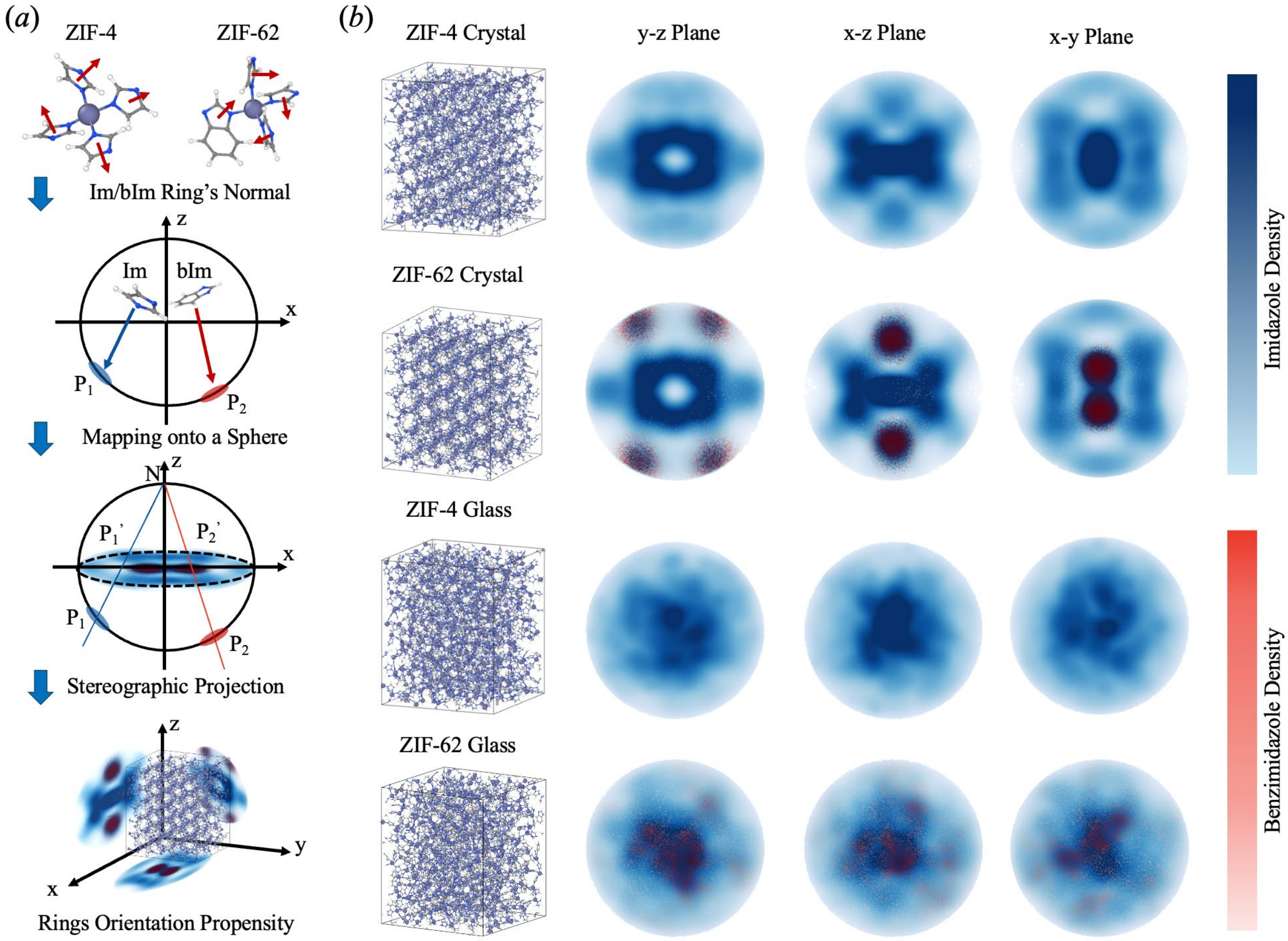

**Figure 1. Orientation propensity of crystalline and glassy ZIF-4/ZIF-62 solid electrolytes.** (*a*) Schematic workflow for quantifying the orientation propensity of Im and bIm rings, shown in blue and red, respectively. Ring normal vectors are extracted onto a unit sphere and mapped onto three orthogonal planes (*y-z*, *x-z*, and *x-y*) using stereographic projection. (*b*) Pole figures showing the resulting orientation of density distribution of Im and bIm rings in crystalline and glassy ZIF-4/ZIF-62, projected onto the three planes.

For crystalline ZIF-62, where bIm rings introduce additional orientational features, the pole figures also display distinct projection-dependent patterns of density distribution as follows. On the *y-z* plane, four localized high-density regions appear at symmetric angular positions of 30°, 150°, 210°, and 330°; on the *x-z* plane, two circular regions emerge at 0° and 180°; and on the *x-y* plane, the elliptical feature observed in ZIF-4 transforms into two adjacent circular regions forming a figure-eight-like pattern. These changes indicate that the larger bIm linkers impose a modified, more complex, yet still well-defined anisotropic orientational distribution.

The fourfold symmetry on the *y*-*z* plane and the twofold symmetry on the *x*-*z* plane are direct fingerprints of the distinct crystallographic environment in the mixed linker framework. In particular, the transformation of the *x*-*y* pole figure into a figure-eight-like pattern should reflect a reorientation of ring normals with direct implications for lithium diffusion along the *z*-direction. Collectively, these orientational preferences define the geometry of ion-transport pathways within the long-range ordered frameworks. As discussed below, the *x*-*y* pole figure, in which a central high-density region indicates preferential alignment of ring normals along the *z*-axis, correlates with the strongly suppressed lithium diffusivity along the *z*-direction in the crystalline phase.

In contrast, the pole figures of glassy ZIF-4 and ZIF-62 show no discrete orientational patterns. Instead, the intensity becomes diffuse on all three projection planes, with a gradual decrease from the center toward the periphery. This centrally concentrated but featureless distribution indicates substantial randomization of ring orientations upon vitrification, while retaining weak orientational memory of the crystalline precursor. Thus, vitrification destroys long-range orientational order, as evidenced by the disappearance of discrete pole-figure features, but preserves short-range orientational correlations reflected in the residual central maximum. The persistence of this central concentration, instead of a perfectly uniform distribution, is expected for a covalently connected network. The ring orientations are not independent variables but constrained by the local Zn-N coordination geometry. Even after long-range order is lost, short-range orientational correlations between neighboring rings remain, giving rise to the observed diffuse but centrally enhanced pattern. From the perspective of ion transport, this near-isotropic orientational distribution of the structural orientation implies that diffusion pathways become homogenized in the glassy state, potentially enabling isotropic ionic conduction. The diffuse pattern is observed for both the Im and bIm rings, indicating that vitrification randomizes the orientations of both linker types despite their different sizes and flexibilities. This similarity further suggests that the underlying vitrification process, driven by local angular distortions, affects the spatial arrangement of Im and bIm linkers in a qualitatively analogous manner. Consequently, the glassy state is expected to exhibit isotropic or near-isotropic lithium diffusion, in contrast to the anisotropic diffusion anticipated for crystalline ZIFs where discrete orientational ordering imposes direction-specific transport constraints.

**2.3 Disorder-enabled low-barrier lithium diffusion.** In ZIF-based materials, ionic conductivity is governed by the collective dynamics of migrating ions, which are highly sensitive to framework topology. We therefore examine how the transition from an ordered crystalline framework to a disordered network affects $Li^+$ transport. The diffusion coefficient ($D$) was obtained from the long-time slope of the mean squared displacement (MSD). As shown in Figure 2*a*, the extrapolated room-temperature diffusion coefficients of crystalline ZIF-4 and ZIF-62 are $1.91 \cdot 10^{-8}$ and $1.12 \cdot 10^{-8}$ $cm^2/s$, respectively. Upon vitrification, these values increase substantially to $3.21 \cdot 10^{-7}$ and $7.76 \cdot 10^{-8}$ $cm^2/s$, respectively. Figure 2*a* further shows that the activation energy decreases from 0.36 and 0.34 eV in crystalline ZIF-4 and ZIF-62, respectively, to 0.16 eV in both glassy phases. This reduction suggests that the disordered framework offers a broader distribution of local environments, allowing $Li^+$ to sample a wider range of migration pathways and bypass the more restrictive, well-defined barriers present in the crystalline lattice. To gain deeper insight into how this disorder alters the microscopic migration behavior, and whether the transport mechanism itself changes beyond a simple barrier reduction, we turn to an analysis of dynamical heterogeneity and displacement distributions.

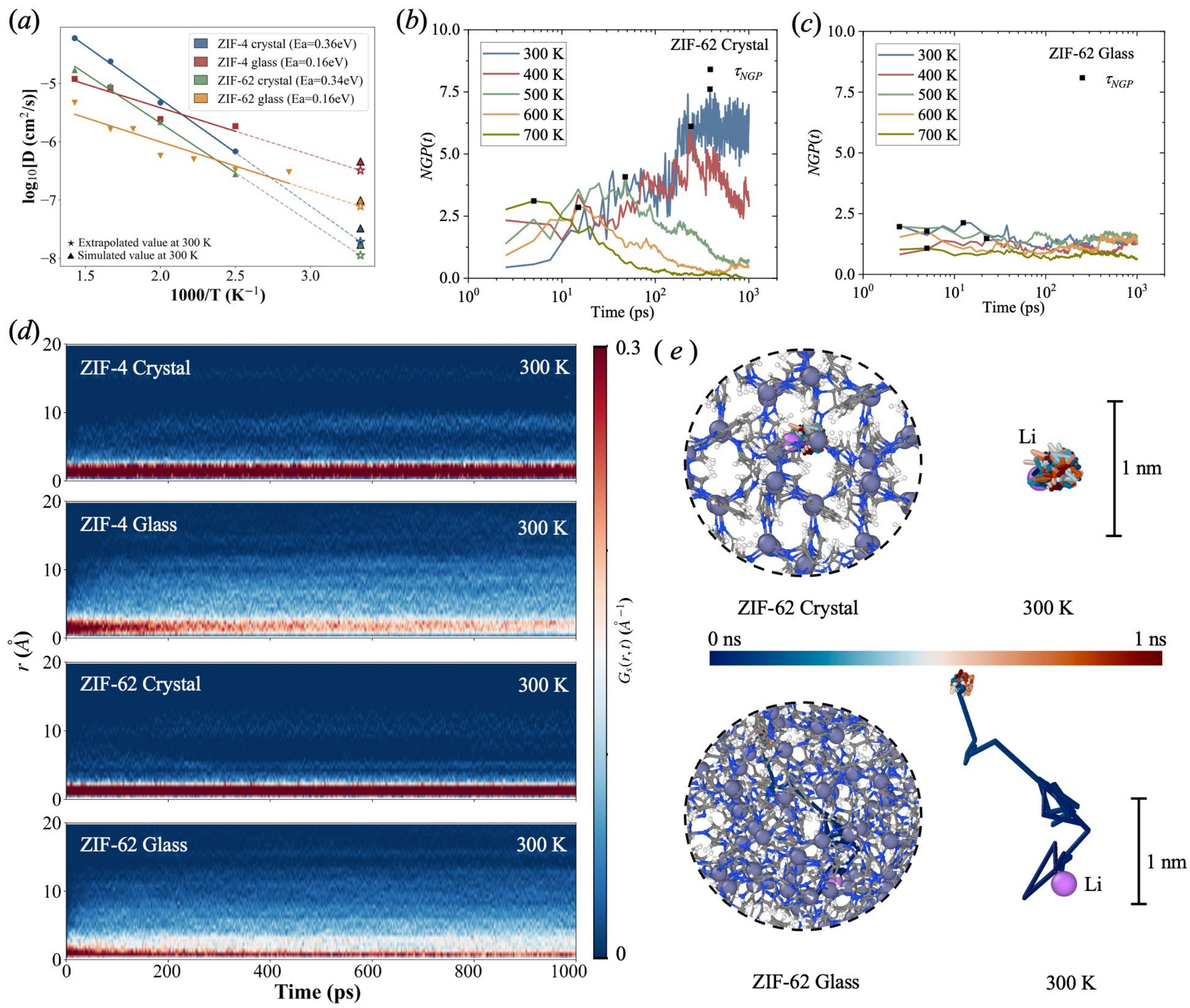

**Figure 2. Disorder-enabled lithium diffusion in crystalline and glassy ZIF electrolytes.** (*a*) Arrhenius analysis of $Li^+$ diffusion coefficients in crystalline and glassy ZIF-4 and ZIF-62. Diffusion coefficients were calculated at 300, 400, 500, 600, and 700 K, with additional simulations at 350, 450, and 550 K for ZIF-62 to improve the reliability of the Arrhenius fits. (*b,c*) Temperature-dependent non-Gaussian parameter (NGP) for $Li^+$ dynamics in (b) crystalline ZIF-4 and (c) glassy ZIF-62. Black squares denote the time at which the NGP reaches its maximum, corresponding to the characteristic timescale for the crossover from dynamically heterogeneous to Gaussian-like diffusion. (*d*) Self-part van Hove correlation function $G_s(r, t)$ for lithium ions in crystalline and glassy ZIF-4/ZIF-62 at room temperature (300 K). (*e*) Representative atomic snapshots comparing $Li^+$ trajectories in crystalline (upper) and glassy (lower) ZIF-62 at 300 K.

To clarify the distinct diffusion mechanisms in the crystalline and glassy phases, we analyze the non-Gaussian parameter (NGP, see Methods section), which is a measure of dynamic heterogeneity through deviations of atomic displacement distributions from Gaussian behavior. A larger NGP indicates more heterogeneous dynamics, where a small fraction of ions undergo rapid hopping events while most remain localized or move slowly. Conversely, a smaller NGP reflects more homogeneous ion motion, approaching Gaussian diffusion. As shown in Figure 2*b* for crystalline ZIF-62 and Figure 2*c* for glassy ZIF-62 (see Supporting Figure S2 for ZIF-4), the NGP evolves differently depending on both structure and temperature. In the crystalline phase, the NGP increases continuously at 300 K within the simulated time window, indicating persistent non-Gaussian dynamics and delayed access to the diffusive regime. At elevated temperatures of 400-700 K, the NGP initially increases to a maximum, and then decreases. The characteristic peak time ($\tau_{NGP}$) shifts to shorter times with increasing temperature, implying that dynamic heterogeneity relaxes more rapidly as thermal activation increases. In contrast, the glassy phase exhibits nearly flat NGP profiles across the investigated temperatures (Figure 2*c*), indicating weak dynamic heterogeneity and more uniform Li-ion motion. Similar trends are observed for crystalline and glassy ZIF-4, as shown in Supporting Figure S2. The $\tau_{NGP}$ value marks the approximate crossover from non-Gaussian to Gaussian dynamics. Beyond this timescale, the system enters the diffusive regime, where the MSD becomes linear with time and $D$ is well defined. This relationship is visualized in Supporting Figure S3, where the MSD curves are shown for crystalline and glassy ZIF-4 and ZIF-62 together with the corresponding $\tau_{NGP}$ values. For crystalline samples at elevated temperatures, the MSD changes from sub-diffusive to diffusive behavior around $\tau_{NGP}$. As temperature decreases, $\tau_{NGP}$ increases and the onset of the diffusive regime is delayed. In contrast, the glassy samples show nearly linear MSD behavior from short times, consistent with homogeneous, Fickian-like diffusion even at low temperature. These results indicate that Li-ion migration in the crystalline frameworks proceeds via preferred pathways dictated by lattice orientation and pore geometry, whereas vitrification introduces a broader distribution of migration channels and reduces the dominance of specific pathways. However, dynamic heterogeneity and ion-transport anisotropy are different phenomena, i.e., to determine whether the glassy phase exhibits isotropic diffusion, the directional diffusion coefficients must be evaluated, which will be done in Section 2.5 below.

Next, to further identify the microscopic origin of the dynamic heterogeneity, we analyze the self-part of the van Hove correlation function $G_s(r,t)$. While the NGP provides an integrated measure of deviations from the Gaussian distribution, $G_s(r,t)$ directly resolves the probability distribution of Li-ion displacements over a given time interval. Therefore, $G_s(r,t)$ can distinguish whether non-Gaussian dynamics arises from broadening of the displacement distribution or from a distinct population of fast-hopping ions. As shown in Figure 2*d*, crystalline ZIF-4 and ZIF-62 exhibit the highest displacement probability near 2 Å at room temperature, indicating that most lithium ions primarily rattle around their initial positions. The ~10 Å feature matches the second Zn-Zn peak, which corresponds to the distance between adjacent cavities. We therefore assign it to a single hop between neighboring cages, while shorter-distance intensity reflects rattling within a single cavity. In contrast, the glassy phases show a broader displacement distribution extending over approximately 0-10 Å, rather than being confined near their initial position. The evolution of $G_s(r,t)$ at 2.5, 10.0, 100, and 1000 ps is shown in Supporting Figure S4. For both glassy ZIF-4 and ZIF-62, the progressive decrease in peak height with time indicates that disorder enables a broader distribution of accessible migration pathways, thereby enhancing room temperature Li-ion diffusion.

Finally, we calculate the non-affine displacement of each $Li^+$ (see Supporting Figure S5 for details). This analysis allows us to classify ions according to their motion in each phase, i.e., those with small non-affine displacements correspond to localized rattling within cages, while those with larger values are associated with hopping events.[31] Based on this classification, we selected typical $Li^+$ from the crystalline and glassy phases to visualize the distinct migration mechanisms. The corresponding trajectories are shown in Figure 2*e*. In the ordered frameworks, lithium ions are largely trapped within voids and undergo localized rattling around equilibrium positions. Long-range ion migration requires escape from well-defined coordination environments and passage through narrow apertures, but these events require overcoming substantial energy barriers, resulting in limited room-temperature diffusion kinetics. By contrast, in the disordered frameworks, $Li^+$ trajectories are smoother and more extended. The disordered network contains a broader spectrum of energy barriers, i.e., some sites impose relatively strong confinement, whereas others provide lower-barrier pathways. As a result, a larger fraction of $Li^+$ becomes mobile even under ambient conditions.

**2.4 Temperature-dependent crossover from hopping to continuous diffusion.** The high-temperature regime of diffusivity allows a clear view of the underlying migration mechanisms as hopping events occur more frequently and statistical sampling is improved. Figure 3*a* shows the self-part van Hove correlation function $G_s(r,t)$ for both crystalline and glassy ZIF-62 at 700 K (see Supporting Figure S6 for corresponding ZIF-4 results). In crystalline ZIF-62, distinct striped feature patterns are observed: as time increases, discrete probability peaks propagate outward along the displacement axis. This stepwise evolution indicates intermittent trapping of lithium ions within framework cavities, followed by rapid stochastic hops to adjacent sites.

Thus, the ordered framework promotes a discrete hopping mechanism, in which Li-ion migration occurs between well-defined coordination environments. In contrast, glassy ZIF-62 exhibits a smoother and more diffuse $G_s(r,t)$ profile, without distinct propagating peaks. The probability density decays gradually with increasing displacement, suggesting that lithium ions in the disordered network experience an energy landscape with a relatively narrow distribution of basin depths. However, the lower diffusivity of the glassy phase at 700 K (Figure 2*a*) may arise from two non-exclusive effects, i.e., transient trapping of ions (as suggested by the persistent features extending to 1000 ps in Figure 3*a*) and a reduced hopping attempt frequency, commonly observed in glasses due to softer vibrational modes.

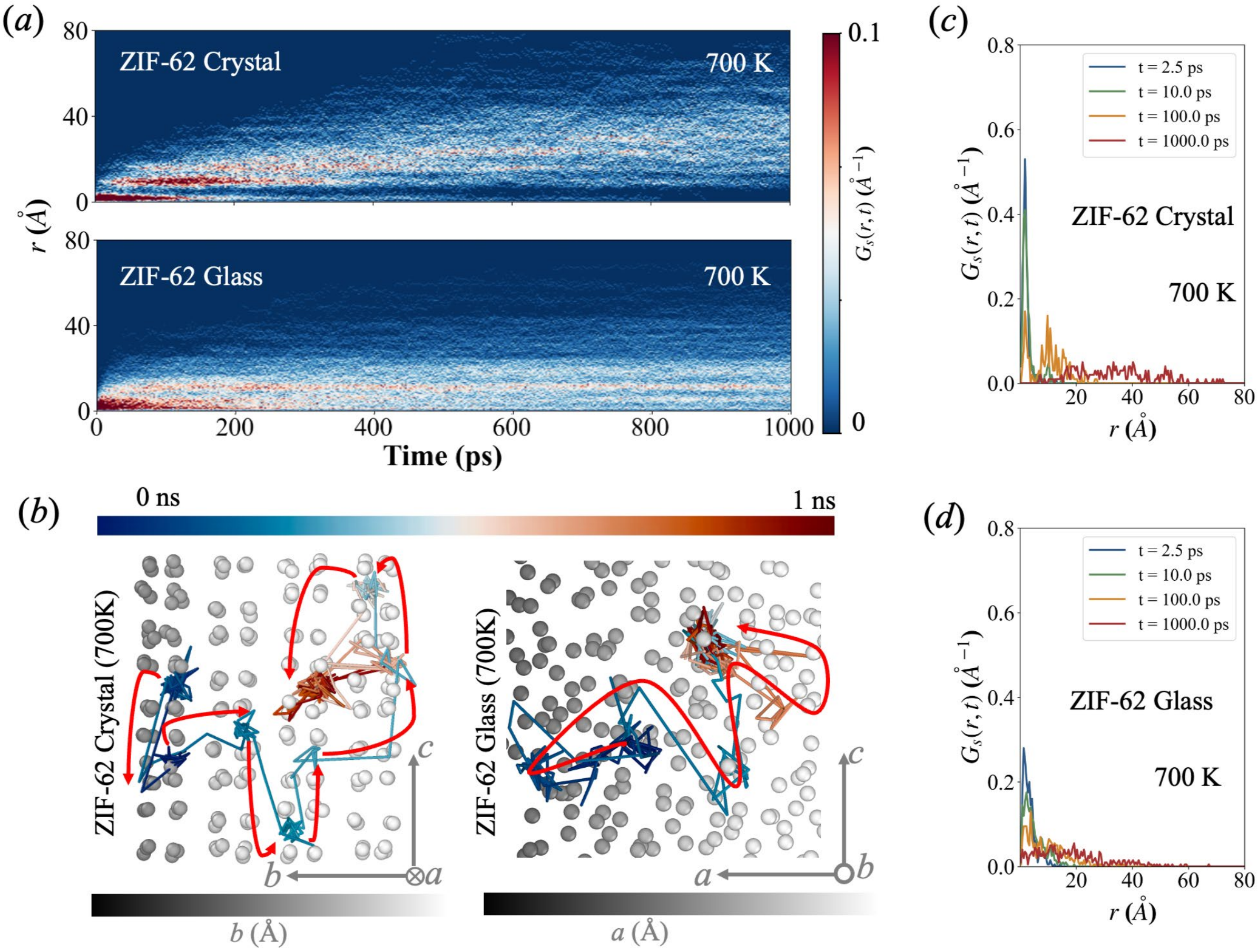

**Figure 3. Transition from discrete hopping to continuous Li-ion migration in ZIF-62.** (*a*) Self-part van Hove correlation function $G_s(r,t)$ for Li ions in crystalline and glassy ZIF-62 at 700 K. (*b*) Schematic illustration of the contrasting $Li^+$ migration mechanisms in crystalline (left) and glassy (right) ZIF-62 at 700 K. In the crystalline framework, Li ions migrate through discrete hopping between well-defined sites, whereas in the glassy framework they undergo smoother, more continuous migration through a disordered energy landscape; gray atoms denote Zn. (*c-d*) Evolution of $G_s(r,t)$ at selected times of 2.5, 10.0, 100, and 1000 ps for (c) ZIF-62 crystal and (d) ZIF-62 glass.

These distinct transport mechanisms are further illustrated schematically in Figure 3*b*. In the crystalline phase, $Li^+$ transport proceeds through discrete jumps between well-defined sites. The rate of this hopping mechanism is strongly temperature-dependent, i.e., at low temperatures, most lithium ions remain trapped because they lack sufficient thermal energy to overcome migration barriers. In contrast, at elevated temperatures, these barriers are more readily overcome, leading to a pronounced increase in diffusivity that ultimately exceeds that of the glassy phase. In the glassy phase, the disordered framework offers a broader distribution of local energy barriers and lower average activation energy, allowing lithium ions to access multiple migration pathways even at room temperature. At high temperature, however, the lack of long-range connectivity of low-energy pathways in the disordered network can restrict overall mobility relative to the crystalline framework, where ordered channels provide continuous percolation paths for fast ion transport once thermal barriers are overcome.

The time evolution of the van Hove function supports this picture (Figures 3*c* and 3*d*). For crystalline ZIF-62 (Figure 3*c*), the probability distribution evolves in a regular, stepwise manner. At the earliest time of 2.5 ps, a single peak near 2 Å dominates, corresponding to localized vibrations of Li ions within their initial cages. As diffusion proceeds and the system approaches the Fickian regime, multiple well-defined peaks emerge. At 100 ps, prominent peaks near 2 Å and 10 Å coexist, indicating that some ions remain confined while others have hopped to neighboring sites. By 1000 ps, the distribution develops multiple fine-structured peaks, reflecting that repeated hopping events are accumulated over the trajectory. In contrast, glassy ZIF-62 (Figure 3*d*) shows no such multiple-peak structure. Instead, the initial peak gradually decreases in intensity and broadens with time, consistent with smoother, more continuous Li-ion migration through the disordered framework.

This mechanistic difference causes a temperature-dependent crossover in diffusion coefficient. Under practical operating conditions for solid-state electrolytes, i.e., near room temperature, the ZIF glasses exhibit substantially higher $Li^+$ diffusion coefficients than their crystalline counterparts (Figure 2*a*). This enhancement arises from the reduced activation energy (~0.16 eV in the glasses versus ~0.35 eV in the crystals), which reflects a broader distribution of accessible migration channels in the disordered framework that increases the population of mobile lithium ions even without strong thermal activation. By contrast, although crystalline ZIFs can exhibit higher diffusivity at elevated temperature once migration barriers are overcome, their room-temperature diffusivity remains limited by $Li^+$ trapping within isolated vacancies. Consequently, for solid-state battery applications under ambient conditions, the glassy ZIF framework offers a pronounced advantage since its energy landscape is more homogeneous with shallow basins, thereby promoting continuous, Fickian-like diffusion.

**2.5 Quantification of anisotropy using diffusion tensor analysis.** To further resolve directional preference of $Li^+$ migration, we analyze the diffusion tensor $D_{ij}$ for crystalline and glassy ZIF-4 and ZIF-62 (see Methods section). The diagonal components $D_{xx}$, $D_{yy}$ and $D_{zz}$ quantify diffusion along the three Cartesian axes, whereas the off-diagonal components $D_{xy}$, $D_{xz}$ and $D_{yz}$ describe cross-correlations between displacementsalong orthogonal directions. In an isotropic system, the diagonal components are equal, i.e., $D_{xx} = D_{yy} = D_{zz}$, and the off-diagonal terms vanish upon ensemble averaging. Anisotropy may therefore arise either from unequal diagonal components, indicating intrinsically direction-dependent diffusion, or from non-zero off-diagonal components, indicating that the principal diffusion axes are not aligned with the chosen Cartesian coordinates or the correlated motion occurs along orthogonal directions. Figure 4*a* illustrates the workflow used to visualize anisotropic diffusion: computing the diffusion tensor from Li-ion trajectories, represented as a diffusivity ellipsoid, mapped onto a logarithmic scale, and projected onto the *y*-*z*, *x*-*z*, and *x*-*y* planes.

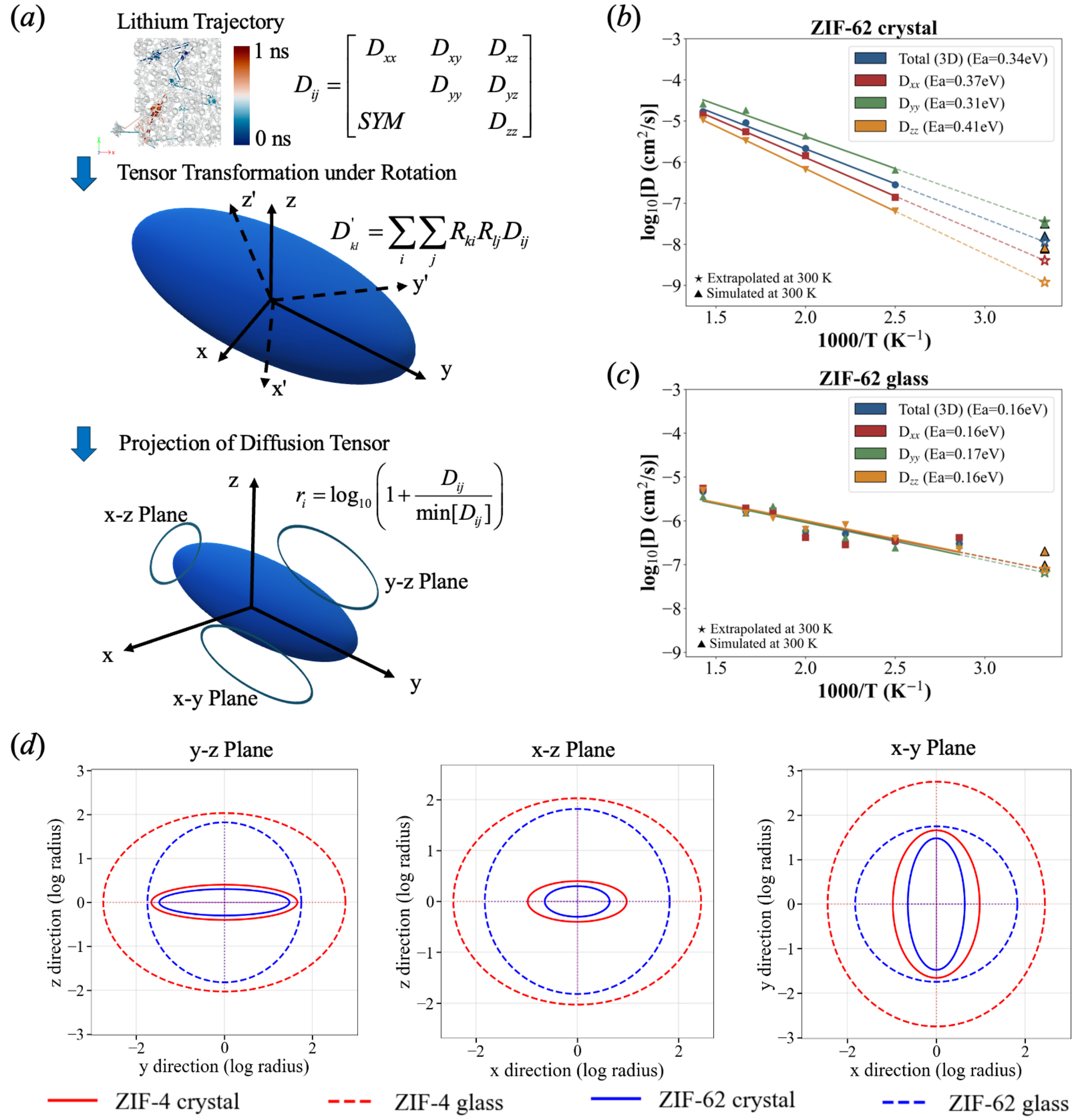

**Figure 4. Disorder-induced transition from anisotropic to isotropic Li-ion diffusion.** (*a*) Workflow for visualizing Li-ion diffusion anisotropy from the diffusion tensor. The tensor is computed from ionic trajectories, represented as a diffusivity ellipsoid, transformed using logarithmic radial scaling, and projected onto three orthogonal planes (*y-z*, *x-z*, and *x-y*). (*b-c*) Temperature-dependent directional diffusion coefficient ($D_{xx}$, $D_{yy}$ and $D_{zz}$), together with the effective three-dimensional diffusion coefficient for (b) crystalline ZIF-62 and (c) glassy ZIF-62. (*d*) Diffusion pole figures projected onto the *y-z*, *x-z*, and *x-y* planes crystalline and glassy ZIF-4/ZIF-62. Red lines denote ZIF-4 and blue lines denote ZIF-62; solid and dashed curves represent crystalline and glassy states, respectively. The evolution from anisotropic, orientation-dependent projections in the crystals to nearly circular isotropic projections in the glasses visualizes the disorder-induced isotropization of $Li^+$ diffusion.

The diagonal components of the diffusion tensor were obtained from Eqs. (6) and (8) (see Methods section) using the directional MSDs ($MSD_{xx}$, $MSD_{yy}$ and $MSD_{zz}$) for crystalline and glassy ZIF-4 and ZIF-62 (see Supporting Figures S7-S8). Although the off-diagonal MSD components also contribute to a complete description of anisotropic diffusion, the calculated cross terms at 300 K (Supporting Figure S9) and 700 K (Supporting Figure S10) are substantially smaller than the diagonal terms for all four systems. The corresponding slopes are close to zero, indicating that the chosen Cartesian axes approximately coincide with the principal axes of the diffusion tensor. We, therefore, approximate $D_{xy}$, $D_{xz}$ and $D_{yz}$ as zero and focus on the diagonal components. These are shown in Figures 4*b* and 4*c* for crystalline and glassy ZIF-62, respectively, together with the total effective three-dimensional diffusion coefficient. The Cartesian axes *x*, *y*, and *z* in our simulations correspond to the crystallographic *a*, *b*, and *c* axes, respectively (see Supporting Figure S11 for the lattice orientation of each sample). In crystalline ZIF-62, Li-ion diffusion is strongly direction dependent, with extrapolated room-temperature diffusion coefficients following the order $D_{yy} > D_{xx} > D_{zz}$. The corresponding activation energies are also diffusion-direction dependent, increasing from 0.31 eV along the *y*-direction to 0.37 eV along *x* and 0.41 eV along *z*. In contrast, glassy ZIF-62 exhibits nearly isotropic transport, with $D_{yy} \approx D_{xx} \approx D_{zz}$ and direction-independent activation energies of approximately 0.16 eV. The $Li^+$ diffusivities for both crystalline and glassy ZIF-4 are provided in Supporting Figure S12 and show the same overall trend, i.e., higher diffusivity, lower activation energy, and more isotropic diffusion in the glassy state relative to the crystalline counterpart. Thus, structural disordering leads to transformation of anisotropic ionic diffusion into isotropic or near-isotropic ionic diffusion transport.

**2.6 Structural origin of lithium diffusion anisotropy.** To identify the structural origin of the anisotropy of $Li^+$ diffusion, we correlate the diffusion pole figures in Figure 4*d* with the ring orientation pole graphs in Figure 1*b*. Since Li-ion migration through the Im and bIm rings is associated with a substantial barrier, the orientation of these rings governs the preferred diffusion pathways. The *x*-*y* projection of the diffusion ellipsoid in Figure 4*d* reflects diffusivity along the *z*-direction. In the corresponding ring-orientation pole figure on the *x*-*y* plane (Figure 1*b*), the central high-density region indicates preferential alignment of ring normals along the *z*-axis. This alignment places ring planes perpendicular to *z*-direction migration and thereby obstructs Li-ion transport along this axis, explaining the suppressed $D_{zz}$ and the oblate shape of the diffusion ellipsoid in the *y*-*z* and *x*-*z* projections. The *y*-*z* projection of the diffusion ellipsoid (Figure 4*d*) primarily reflects diffusivity along the *x*-direction. In the corresponding ring orientation pole figure (Figure 1*b*), the hollow, curved-square central region indicates that the ring normals deviate only slightly from the *x*-direction, leaving comparatively accessible pathways for Li migration along *x*. For crystalline ZIF-62, the *x*-*z* pole figure further shows a dog-bone-shaped distribution of Im-ring orientations centered along the horizontal direction (90°-270°), indicating that the Im ring normals predominantly lie near the *x*-axis. As a result, the Im ring planes tend to avoid blocking the *y*-direction, preserving open channels for Li-ion migration along *y*. Although bIm rings introduce additional features at 0° and 180°, suggesting local obstruction of *y*-direction diffusion, their low concentration means that these Im/bIm-induced local barriers do not dominate the global transport behavior. Consequently, *y*-direction diffusion remains fastest, whereas *x*-direction diffusion is intermediate, and *z*-direction diffusion is most strongly suppressed.

In the glassy state, the pole figures of ZIF-4 and ZIF-62 show no discrete orientational features (Figure 1*b*). Instead, the intensity becomes diffuse across all three projection planes, indicating substantial randomization of Im and bIm ring orientations. As a result, the ring planes no longer impose a directional bias on Li-ion migration, and pathways that are obstructed in the crystalline phase become accessible. The diffusion pole graphs in Figure 4*d* directly support this interpretation. For ZIF-62 glass, the *y*-*z*, *x*-*z*, and *x*-*y* planes projections are nearly circular, demonstrating essentially isotropic Li-ion diffusion. ZIF-4 glass retains a weak residual anisotropy, as indicated by a slightly elongated *x*-*y* projection and by the ordering $D_{yy}$ >

$D_{xx} > D_{zz}$ in Supporting Figure S13, although the differences in $Li^+$ diffusivity are smaller compared to its crystalline counterpart. This residual anisotropy likely reflects partial orientational memory of the crystalline precursor, consistent with the centrally concentrated diffuse pattern in the orientation pole figures. In ZIF-62 glass, the presence of bIm linkers further disrupts orientational correlations during melt-quenching, leading to more complete isotropization of Li-ion transport.

Collectively, these results establish a direct link between linker-ring orientation and Li-ion diffusion anisotropy. Ordered ring alignment in crystalline ZIFs creates direction-specific migration and produces anisotropic diffusion ($D_{yy} > D_{xx} > D_{zz}$), whereas orientational disorder in the glass removes most directional constraints and enables isotropic or near-isotropic diffusion. Together with the reduced energy barriers shown in Figure 2*a*, the anisotropy-to-isotropy transition provides a complete picture of the transport benefits of vitrification, i.e., lower barriers and homogenized pathways both contribute to the enhanced room-temperature $Li^+$ mobility in ZIF glasses.

## 3. CONCLUSION

This study establishes that framework disorder and linker-ring orientation are key structural descriptors governing the kinetics and anisotropy of $Li^+$ diffusion in ZIF-based solid electrolytes. Using a machine learning interatomic potential fine-tuned to include lithium interactions, we systematically investigated lithium ion transport in crystalline and glassy ZIF-4 and ZIF-62. Compared with their crystalline counterparts, the glassy phases exhibit markedly reduced activation energies, from approximately 0.35 to 0.16 eV, and substantially enhanced extrapolated room-temperature diffusion coefficients. The structural disordering induced by vitrification increases the $Li^+$ diffusion coefficient by more than an order of magnitude in ZIF-4, from $1.91 \cdot 10^{-8}$ to $3.21 \cdot 10^{-7}$ $cm^2/s$, and by nearly seven times in ZIF-62, from $1.12 \cdot 10^{-8}$ to $7.76 \cdot 10^{-8}$ $cm^2/s$. In addition, the structural disordering alters the $Li^+$ migration mechanism. Crystalline ZIFs exhibit pronounced dynamic heterogeneity, with lithium ions undergoing rare hopping events between well-defined framework cavities. In contrast, ZIF glasses show more homogeneous, Fickian-like diffusion through a broadened distribution of local environments. At the microscopic level, the preferred orientation of Im and bIm rings imposes energy barriers for direction-specific migration barriers in the crystals, leading to strongly anisotropic diffusion. Upon vitrification, these ring orientations become randomized, removing most directional constraints and transforming anisotropic $Li^+$ transport into isotropic or near-isotropic diffusion. These findings reveal a direct structure versus ion transport relationship, i.e., the link between framework disorder, linker-ring orientation, activation barriers, and $Li^+$ mobility. ZIF glasses therefore provide a promising design platform for developing isotropic, room-temperature solid electrolytes for next-generation all-solid-state batteries.

## 4. METHODS

**4.1 Machine learning interatomic potential.** Previous studies have demonstrated that the DeePMD force field used for ZIF glasses provides near density functional theory (DFT) accuracy and reproduces available experimental observations.[23, 24] This force field was originally trained using the DeePMD-kit code framework[21, 22] on a DFT dataset. This dataset involves C, H, N, and Zn elements, spanning crystalline and amorphous configurations of representative ZIF systems, including ZIF-4, SALEM-2, ZIF-8, ZIF-62, together with relevant elemental and compound reference structures. Since the original model did not include lithium interactions, we extended the training dataset by incorporating configurations from a liquid electrolyte/ZIF interface model, designed to capture the solvation environment and interfacial $Li^+$ behavior relevant to experimental battery conditions. In this model, the liquid electrolyte consists of four ethylene carbonate (EC), four dimethyl carbonate (DMC), four diethyl carbonate (DEC), two $LiPF_6$, and one vinylene carbonate (VC) molecules, placed in contact with the ZIF-based solid electrolyte. The original ZIF potential was then fine-tuned on this augmented dataset using the same DeePMD-kit workflow.[21, 22] Notably, lithium species are consistently treated as $Li^+$ in all simulations, in line with their treatment in the DFT training data.

To validate the transferability and accuracy of the fine-tuned model beyond the training distribution, we constructed a separate test set based on a direct $Li^+$ insertion model, where $Li^+$ ions were placed at interstitial sites within the ZIF frameworks. In the *ab initio* MD calculations used to generate this test set, the ZIF framework was treated as neutral and each inserted Li was assigned a formal charge of +1. This setup explicitly evaluated the model's ability to capture $Li^+$-framework interactions and migration barriers within the ZIF bulk structure. The performance of the DeePMD model was then benchmarked against DFT calculations on this test set. The resulting root-mean-square errors were $5.51\cdot10^{-3}$ eV/atom for energy and $1.12\cdot10^{-1}$ ev/Å for force, as summarized in Supporting Figure S12, confirming that the model maintains DFT-level accuracy even for configurations outside the original training distribution.

**4.2 ZIF crystal/glass sample preparation.** The initial crystallographic information files (CIFs) for the ZIF-4 and ZIF-62 structures were obtained from the Cambridge Structural Database (CSD).[24, 32] Both frameworks were extensively studied in previous literature.[24, 29] We note that multiple compositions exist for ZIF-62, as the molar ratio between benzimidazolate (bIm) and imidazolate (Im) can vary. In this work, we adopted the composition $ZnIm_{1.75}bIm_{0.25}$ as a representative example. A supercell expansion of 3×3×3 replicas was applied to the conventional unit cell obtained from the CIF, resulting in 7,344 atoms for ZIF-4 crystal/glass and 7,992 atoms for ZIF-62 crystal/glass. This resulted in a simulation box containing enough atoms for subsequent MD simulations.

All MD runs were performed using LAMMPS with a DeePMD-kit potential. The force field parameters covered the element types C, H, N, Zn and Li. A timestep of 0.25 fs was used throughout, with periodic boundary conditions applied in all three directions. The neighbor list cutoff was set to 1.0 Å beyond the pairwise interaction range. To obtain the crystalline states of ZIF-4 and ZIF-62 for subsequent lithium diffusion coefficient calculations, the structures were subjected to a relaxation protocol at 300 K. This protocol consisted of a 12.5 ps equilibration under *NPT* conditions, followed by a 25 ps production run under *NVT* dynamics.

To obtain the glassy states of ZIF-4 and ZIF-62, a two-stage melt-quenching protocol was employed. A direct single-step quenching applied to the 3×3×3 supercell led to unrealistically low densities compared to experimental measurements. Therefore, a two-step strategy was implemented to achieve denser and more thoroughly relaxed glass structures.[24] In the first stage, the 3×3×3 supercell underwent an initial *NPT* equilibration at 10 K for 7.5 ps using a Berendsen thermostat, after which the temperature was raised to 300 K over 1.25 ps. The system was then heated from 300 K to 1500 K over 50 ps and maintained at 1500 K for another 50 ps under a constant applied pressure of 1 GPa. This high-pressure condition ensured complete melting while preserving the integrity of the organic linkers. The melt was subsequently cooled from 1500 K down to 300 K at a rate of 5 K/ps (total quenching time: 240 ps), still under 1 GPa. The product of this stage is referred to as the as-quenched glass.

In the second stage, the as-quenched glass was further processed under ambient pressure to eliminate the residual internal stress remaining from the first stage. The sample was again heated from 300 K to 1500 K over 50 ps, held at 1500 K for 50 ps, and then cooled back to 300 K at the same rate of 5 K/ps (240 ps). A final relaxation step was performed at 300 K, with 12.5 ps under *NPT* conditions, followed by 25 ps of *NVT* dynamics for production runs. Atomic configurations were saved every 0.25 ps, i.e., every 1000 timesteps, for subsequent structural analysis. The exact same two-stage protocol was applied to ZIF-62, using identical temperature (1500 K) and pressure settings (1 GPa in first stage and 0 GPa in second two).

**4.3 Structural descriptors.** To characterize the local atomic arrangements in the simulated structures of ZIF-4 and ZIF-62, we calculated the partial pair distribution function (PDF), $g_{\alpha\beta}(r)$, which describes the probability of finding a $\beta$ atom at a distance $r$ from an $\alpha$ atom. Its definition is given by,

$$g_{\alpha\beta}(r) = \frac{1}{\rho_\beta} \cdot \frac{dn_{\alpha\beta}(r)}{4\pi r^2 dr}, \tag{1}$$

where $dn_{\alpha\beta}$(r) counts the number of $\beta$ atoms located within a spherical shell of thickness $dr$ at radius $r$ around $\alpha$ atoms, and $\rho_\beta$ denotes the number density of $\beta$ atoms. The total PDF, $g(r)$, is then obtained as a sum of all partial terms as,

$$g(r) = \sum_\alpha \sum_\beta g_{\alpha\beta}(r). \quad (2)$$

Partial structure factors $S_{\alpha\beta}(Q)$ of the ZIF crystal/glass were calculated by using the Fourier transformation of $g_{\alpha\beta}(r)$ using the Faber-Ziman formalism, which is expressed as,

$$S_{\alpha\beta}(Q) = 1 + 4\pi\rho \int_0^{r_{\max}} r^2 \frac{\sin(Qr)}{Qr} \left[ g_{\alpha\beta}(r) - 1 \right] dr, \quad (3)$$

where $q$ is the magnitude of the scattering vector, $\rho$ is the overall atomic number density, and $r_{\max}$ is the upper integration limit, typically taken as half the side length of the simulation box. Based on Eq. (2), the neutron-weighted and X-ray weighted total structural factor $S_{neutron}(Q)$ and $S_{X\text{-}ray}(Q)$ were calculated as,

$$S(Q) = \sum_\alpha \sum_\beta c_\alpha c_\beta b_\alpha b_\beta S_{\alpha\beta}(Q), \quad (4)$$

where $c_\alpha$ and $c_\beta$ are the atomic fractions of species $\alpha$ and $\beta$, while $b_\alpha$ and $b_\beta$ represents the neutron/X-ray scattering lengths of the corresponding nuclei. For the elements present in ZIF-4 and ZIF-62, the neutron scattering lengths were taken as 6.646 fm, -3.739 fm, 9.36 fm, 5.68 fm, and -3.739 fm for C, H, N, Zn, and Li, respectively, while the X-ray scattering lengths were taken as 6.0 e.u., 1.0 e.u., 7.0 e.u., 30.0 e.u., and 3.0 e.u. for C, H, N, Zn, and Li, respectively.

**4.4 Rings orientation descriptors.** To quantify the medium-range structural order in ZIF-4 and ZIF-62, we adopted the ring orientation descriptor, originally proposed for imidazole rings in ZIF-4.[33] In the present work, we extended this descriptor to benzimidazole linkers in ZIF-62, where both the imidazole ring and the fused benzene ring were considered separately. For each imidazole ring, the ring plane normal vector was calculated from the cross product of two non-parallel vectors connecting the ring atoms. For each benzene ring in benzimidazole, the normal vector was defined similarly using its six carbon atoms. All normal vectors were then normalized to unit vectors. Stereographic projections of these unit vectors onto the *yz*-, *xz*-, and

*xy*-planes (i.e., $x$=0, $y$=0, and $z$=0) were performed to generate pole figures. The orientation propensity was characterized by the density distribution of the projected points on each plane. A concentrated pole figure indicates a preferred ring orientation (e.g., in the crystalline state), whereas a more uniform pole figure reflects higher orientational disorder (e.g., in the glass or liquid state). By comparing the pole figures of imidazole and benzene rings in ZIF-62, we distinguished the distinct roles of the two ring types in the crystals and glasses.

**4.5 Lithium-ion diffusion simulations in ZIF-4 and ZIF-62.** Lithium diffusion in ZIF-4 and ZIF-62 frameworks was determined using mean squared displacement (MSD) analysis from the MD trajectories. Before MSD evaluation, each Li-loaded ZIF structure was relaxed under *NVT* conditions at several target temperatures, with a total simulation length of 1 ns. In this work, we focused on the dilute solution regime, where the lithium concentration was kept low. Accordingly, only 10 Li ions were introduced into each 3×3×3 supercell of the ZIF-4 (7,344 atoms) and ZIF-62 (7,992 atoms) frameworks. However, such a small number of Li ions per simulation box was insufficient to achieve statistically meaningful convergence of the MSD and the derived $Li^+$ diffusion coefficient. To overcome this limitation and enhance statistical sampling, we adopted a replica-based approach. Specifically, we generated 20 independent replicas for each crystalline ZIF system and 40 independent replicas for each glassy ZIF system. These replicas differed in both the random initial velocities assigned to all atoms and the random initial positions assigned to the lithium ions. For each replica, a 1 ns MD production run was performed. The final MSD was then computed as an ensemble average over all Li ions across all replicas, yielding sufficiently converged statistics for subsequent diffusion analysis. Throughout these simulations, lithium is treated as $Li^+$, consistent with the charged species represented in the DFT training data (see Section 4.1). The total MSD was defined as the ensemble-average squared displacement of lithium ions as,

$$\mathrm{MSD}(t) = \frac{1}{N}\sum_{i=1}^{N}\left[\vec{r}_i(t) - \vec{r}_i(0)\right]^2, \quad (5)$$

where $\vec{r}_i(t)$ and $\vec{r}_i(0)$ are the position vector of the $i$-th Li ion at time $t$ and at initial position, respectively, and $N$ is the total Li ion number in the simulation cell. For anisotropic diffusion behavior, the MSD along the $x$, $y$ and $z$ directions, including that cross term of the MSD, were defined as,

$$\mathrm{MSD}_{\alpha\beta}(t)=\frac{1}{N}\sum_{i=1}^{N}\left[\vec{r}_{i,\alpha}(t)-\vec{r}_{i,\alpha}(0)\right]\left[\vec{r}_{i,\beta}(t)-\vec{r}_{i,\beta}(0)\right], \tag{6}$$

where $\alpha,\beta=x,y,z$, and thus $\vec{r}_{i,x}(t)$, $\vec{r}_{i,y}(t)$ and $\vec{r}_{i,z}(t)$ are the position $x$, $y$ and $z$ components of the position vector at time $t$, while $\vec{r}_{i,x}(0)$, $\vec{r}_{i,y}(0)$ and $\vec{r}_{i,z}(0)$ are the position $x$, $y$ and $z$ components of the position vector at the initial position. When $\alpha=\beta$, $\mathrm{MSD}_{\alpha\alpha}(t)$ represents the ensemble-average squared displacement of lithium ions along $x$, $y$ and $z$ directions; while for $\alpha\neq\beta$, $\mathrm{MSD}_{\alpha\beta}(t)$ represents the cross term that measures the cross-correlations between displacements along orthogonal directions. For example, $\mathrm{MSD}_{xy}(t)$ quantifies the coupling between $x$ and $y$ displacements. In an isotropic system such ZIF glass, these cross terms vanish upon ensemble averaging, indicating that diffusive motions along different Cartesian axes are statistically independent. Conversely, non-zero values indicate anisotropic diffusion or a misalignment between the coordinate axes and the principal axes of the diffusion tensor. For each simulated ZIF structure in three dimensions, the self-diffusion coefficient $D$ of lithium was obtained from the long-time slope of the MSD curve,

$$D=\frac{1}{6}\lim_{t\to\infty}\frac{d}{dt}\mathrm{MSD}(t). \tag{7}$$

Based on Eq. (6), the directional diffusion coefficients and the cross term of the diffusion coefficient could be expressed as,

$$D_{\alpha\beta}=\frac{1}{2}\lim_{t\to\infty}\frac{d}{dt}\mathrm{MSD}_{\alpha\beta}(t). \tag{8}$$

Here, $\alpha=\beta$ represents the diffusion coefficient along $x$, $y$, and $z$ directions; while $\alpha\neq\beta$ represents the cross term that measures the cross-correlations between diffusivity along orthogonal directions. The anisotropic diffusion behavior allowed the $Li^+$ diffusion along the $x$, $y$, and $z$ directions with diffusion coefficient, where the diffusion tensor is expressed as,[34]

$$D_{ij} = \begin{bmatrix} D_{xx} & D_{xy} & D_{xz} \\ & D_{yy} & D_{yz} \\ SYM & & D_{zz} \end{bmatrix}, \tag{9}$$

where $i, j = x, y$ and $z$. For the diagonal tensor in Eq. (9), $D_{xx}$, $D_{yy}$ and $D_{zz}$ are the principal diffusion direction, but if $D_{xy}$, $D_{xz}$ and $D_{yz}$ are not equal to zero, the principal diffusion direction can be determined by using the tensor transformation under rotation as,

$$D'_{kl} = \sum_i \sum_j R_{ki} R_{lj} D_{ij} , \tag{10}$$

where the rotation matrix $R_{ij}$ is defined based on the rotation angle about the $x$-, $y$- and $z$-axis. By using the x-y-z Euler angle convection, the net rotation matrix is expressed as,

$$R_{ik} = \sum_j \sum_l R_{ij}^z(\varphi) R_{jl}^y(\phi) R_{lk}^x(\theta), \tag{11}$$

where $R_{ij}^z(\varphi)$, $R_{ij}^y(\phi)$ and $R_{ij}^x(\theta)$ are defined by the rotation angles $\varphi$, $\phi$, and $\theta$ as,

$$R_{ij}^z(\varphi) = \begin{bmatrix} 1 & 0 & 0 \\ 0 & \cos\varphi & \sin\varphi \\ 0 & -\sin\varphi & \cos\varphi \end{bmatrix}, \tag{12}$$

$$R_{jl}^y(\phi) = \begin{bmatrix} \cos\phi & 0 & -\sin\phi \\ 0 & 1 & 0 \\ \sin\phi & 0 & \cos\phi \end{bmatrix}, \tag{13}$$

$$R_{lk}^x(\theta) = \begin{bmatrix} \cos\theta & \sin\theta & 0 \\ -\sin\theta & \cos\theta & 0 \\ 0 & 0 & 1 \end{bmatrix}, \tag{14}$$

respectively. By using Eqs. (10)-(14), the diffusion coefficient along an arbitrary direction can be calculated, and anisotropic diffusion tensor can be illustrated as a ellipsoid plane while the isotropic diffusion tensor can be illustrated as a spherical plane. The diffusion tensor $D_{ij}$ defines an ellipsoid whose semi-axis lengths are proportional to $\sqrt{D_{ij}}$ . This ellipsoid represents the directional dependence of molecular mobility. To analyze the anisotropy in different crystallographic planes, the 3D ellipsoid was projected onto the $x$-$y$, $x$-$z$ and $y$-$z$ planes, yielding ellipses that directly reveal the diffusional anisotropy in each plane. However, when comparing different materials, the absolute values of $D_{ij}$ often differ by several orders of magnitude. Typically, the glassy ZIF-4 and ZIF-62 exhibit diffusion coefficients that are 10-100 times larger than those of the crystalline ZIF-4 and ZIF-62. Plotting the ellipses with linear scaling would shrink the crystal ellipses to nearly invisible points. To overcome these issues while preserving the relative anisotropy information, we employed a logarithmic radial transformation. For each principal direction $i$, the effective radius used for plotting was defined as,

$$r_i = \log_{10}\left(1+\frac{D_{ii}}{D_{\min}}\right),\ \ D_{\min} = \min_{all\ system,\, k\in\{x,\,y,\,z\}} D_{kk}\ , \tag{15}$$

where $D_{\min}$ is the smallest diffusion coefficient among crystalline and glassy ZIF-4 and ZIF-62. This transformation compressed the wide numerical range while retaining the ability to compare shape. For example, an isotropic material produces a circular projection, whereas anisotropic appear with comparable ellipse sizes, allowing a clear visual assessment of how anisotropy changes upon vitrification.

The temperature dependent diffusion coefficient $D$ was then analyzed by fitting to the Arrhenius equation,

$$D(T) = D_0 \exp\left(\frac{-E_a}{k_B T}\right), \tag{16}$$

where $E_a$ is the activation energy for Li migration, $k_B$ is the Boltzmann constant, $T$ is the absolute temperature, and $D_0$ is the pre-exponential factor corresponding to the diffusivity at infinite temperature.

To measure the deviation of particle displacement magnitudes distribution from a Gaussian distribution, which is also referred to as the dynamic heterogeneity, we calculated the so-called Non-Gaussian parameter (NGP) as,

$$NGP(t) = \frac{3}{5}\frac{\frac{1}{N}\sum_{i=1}^{N}\left[\vec{r}_i(t)-\vec{r}_i(0)\right]^4}{\left\{\frac{1}{N}\sum_{i=1}^{N}\left[\vec{r}_i(t)-\vec{r}_i(0)\right]^2\right\}^2} - 1\,. \tag{17}$$

NPG has a minimal theoretical value of -0.4, for which the displacement magnitude of all atoms is the same. If it is equal to 0, it is equivalent to Brownian diffusion, and if NPG > 0, some mobile atoms would move faster than others.

To measure the detailed ionic migration pathways beyond the average behavior captured by the mean squared displacement, we also introduced the self-part of the van Hove. This function quantifies the probability that a Li-ion has moved a distance $r$ after a time $t$, thereby revealing the underling diffusion mechanism, including continuous, hopping, or heterogeneous transport, in glassy ZIF electrolytes. The self-part of the van Hove is defined as,

$$G_s(r,t) = \frac{1}{N}\sum_{i=1}^{N}\left[\delta\left(r - \left|\vec{r}_i(t) - \vec{r}_i(0)\right|\right)\right]. \tag{18}$$

where $r$ denotes the displacement magnitude.

**ACKNOWLEDGEMENTS**

This work was supported by a MSCA Postdoctoral Fellowship (101148843) from Horizon Europe. T.J. and K.T. acknowledge support from the Novo Nordisk Foundation (NNF23OC0087524)

**CODE AND DATA AVAILABILITY**

The complete simulation workflows, including the generation of ZIF glass structures and MD run protocols, together with the trained deep learning potential, are available on GitHub (https://github.com/clyongAAU/AAU_glass_ZIF). All source data supporting the conclusions of this work are provided in the paper and its supplementary information.

# Supporting Information

*for*

## Fast Isotropic Li-Ion Diffusion in Zeolitic Imidazolate Framework Glass Electrolytes for Batteries

Yong Li [a], Tao Du [b,*], Timothée Jamin [a], Zhencai Li [a], Kasper Tolborg [a], Yuanzheng Yue [a], Morten M. Smedskjaer [a,*]

[a] *Department of Chemistry and Bioscience, Aalborg University, 9220 Aalborg East, Denmark*

[b] *Department of Applied Physics, The Hong Kong Polytechnic University, Kowloon, Hong Kong 999077, China*

[*] *Corresponding authors. E-mail: mos@bio.aau.dk (M.M.S.); dutaohit@gmail.com (T.D.)*

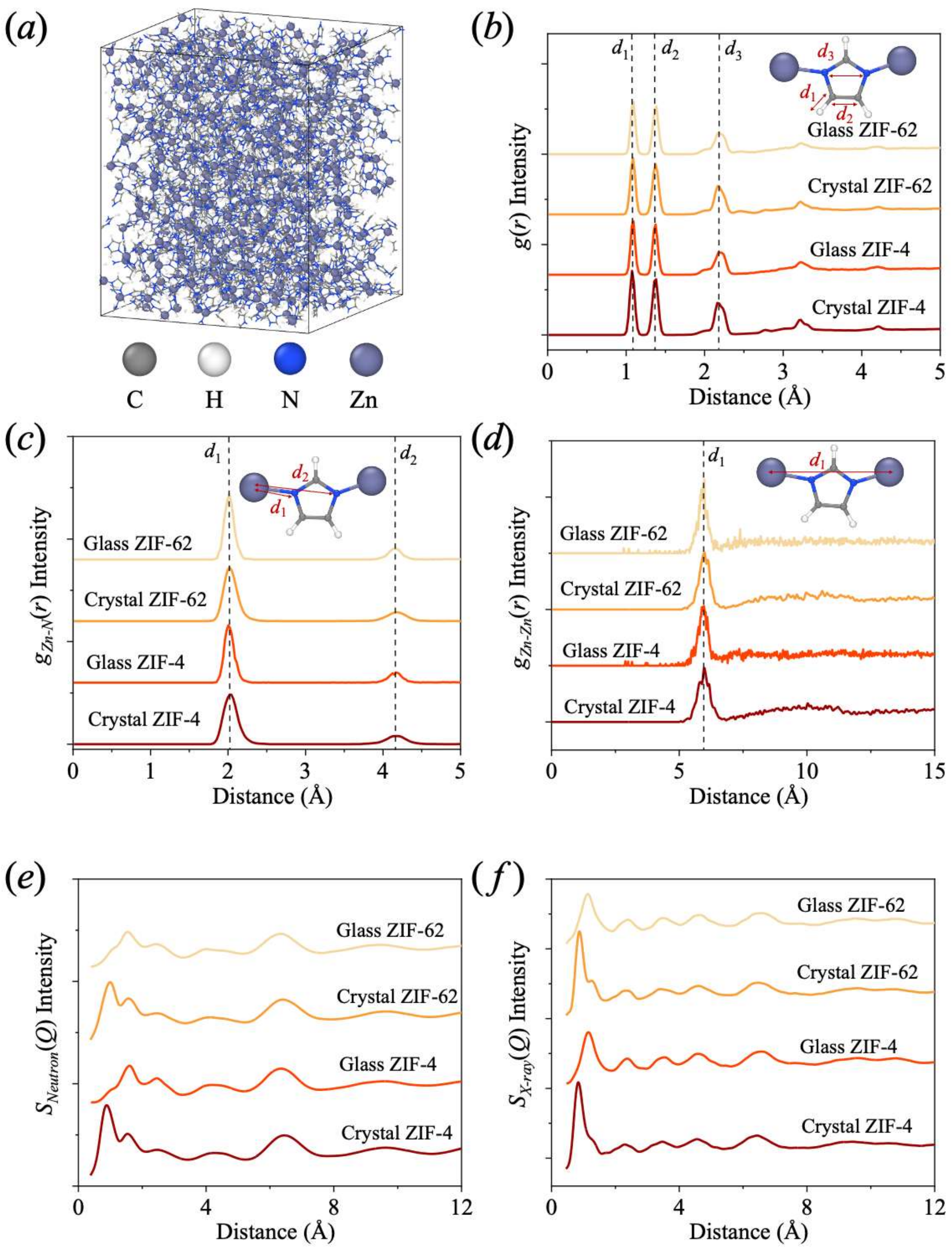

**Figure S1.** Structural fingerprints of the ZIF structural transition from ordered to disordered states. (*a*) Atomic snapshot of a simulated ZIF structure, showing C, H, N, and Zn atoms. (*b*) Total pair distribution function $g(r)$ of crystalline and glassy ZIF-4 and ZIF-62, simulated by MD using a machine learning interatomic potential. (*c*) Pair distribution function for Zn-N bonds, reflecting the local coordination environment around Zn. (*d*) Pair distribution function for Zn-Zn pairs, sensitive to medium-range ordering. (*e*) Simulated neutron scattering structure factor $S_{Neutron}(Q)$ for ZIF crystal and glass. (*f*) Simulated X-ray scattering structure factor $S_{X\text{-}ray}(Q)$ for ZIF crystal and glass.

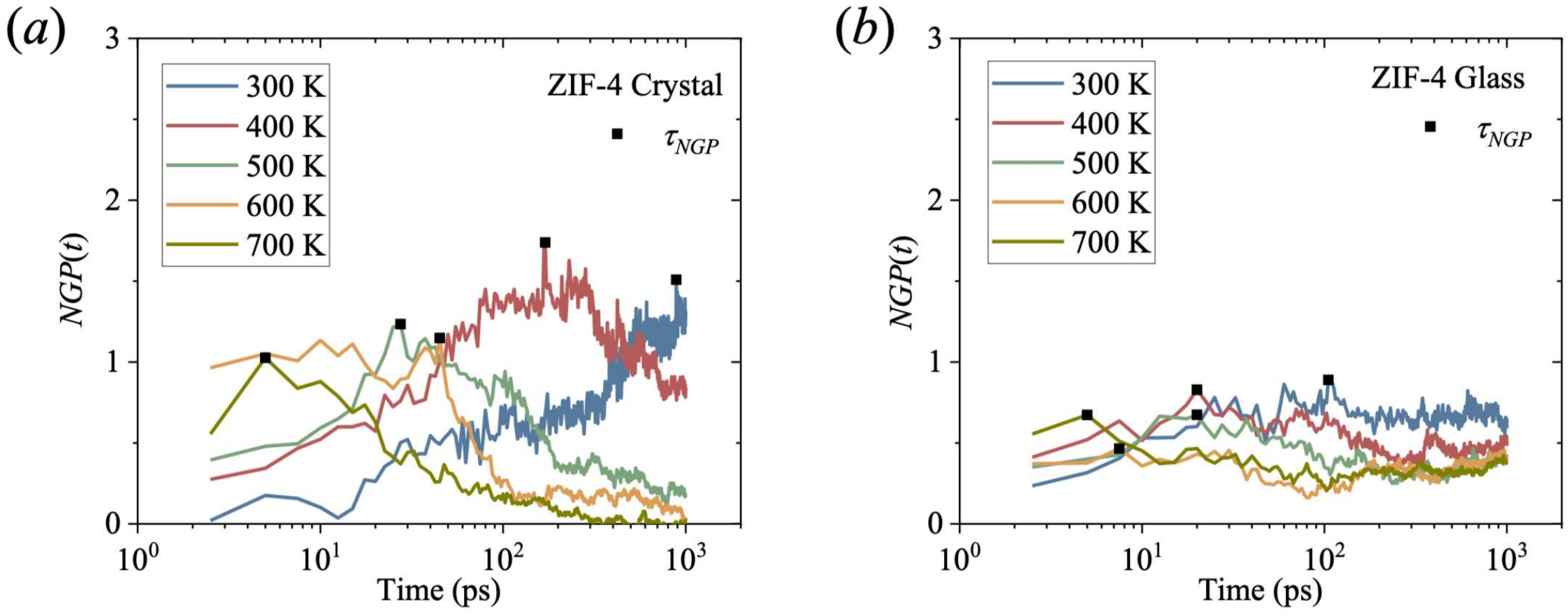

**Figure S2.** Non-Gaussian parameter (NGP) of lithium ions at different temperatures for (*a*) crystalline ZIF-4 and (*b*) glassy ZIF-4. The black squares denote the time at which the NGP reaches its maximum value, indicating the onset of dynamical heterogeneity.

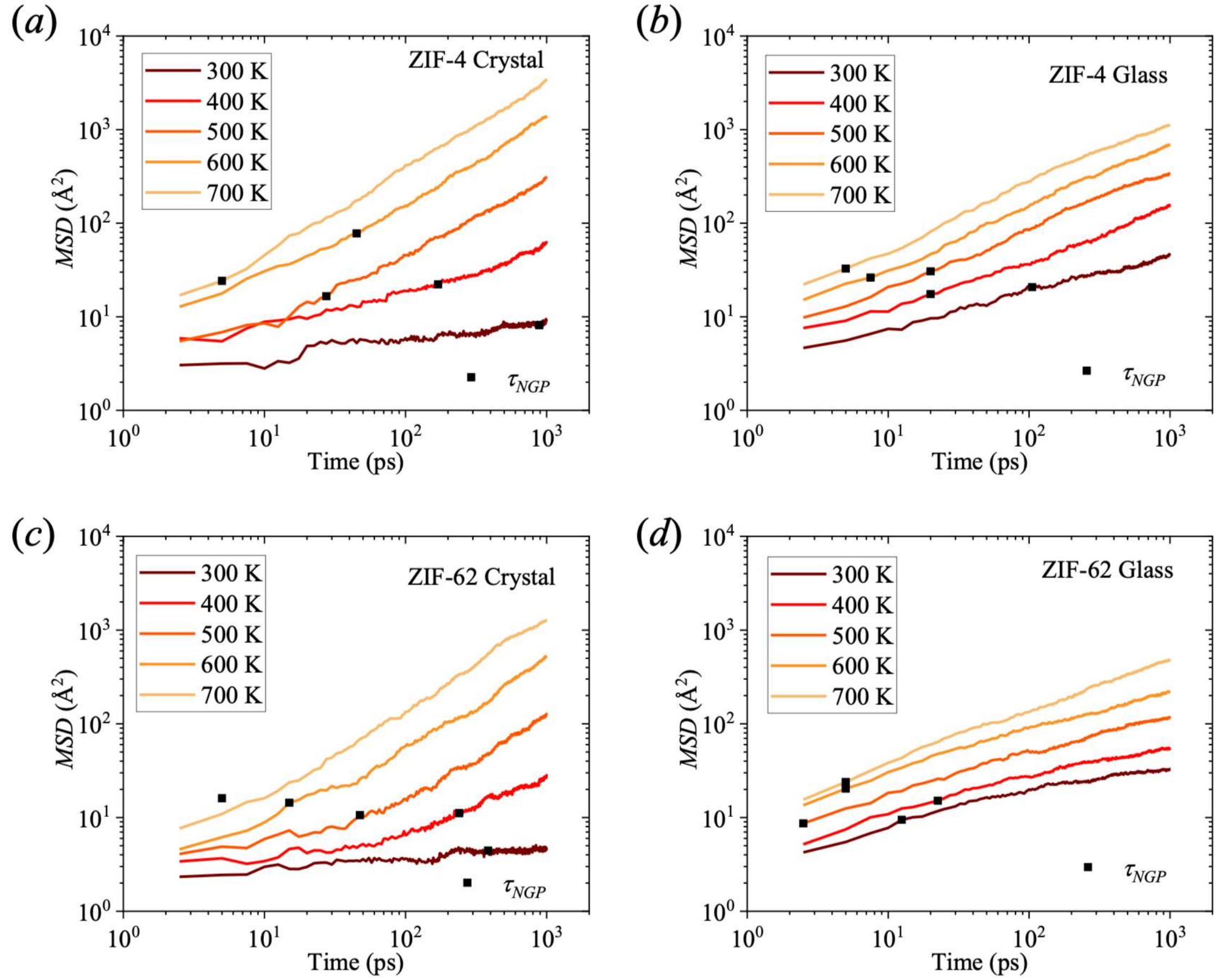

**Figure S3.** Mean squared displacement (MSD) of lithium-ions in (*a*) crystalline ZIF-4, (*b*) glassy ZIF-4, (*c*) crystalline ZIF-62, and (*d*) glassy ZIF-62 at different temperatures. The black squares denote the time at which the NGP reaches its maximum value, indicating the onset of dynamical heterogeneity.

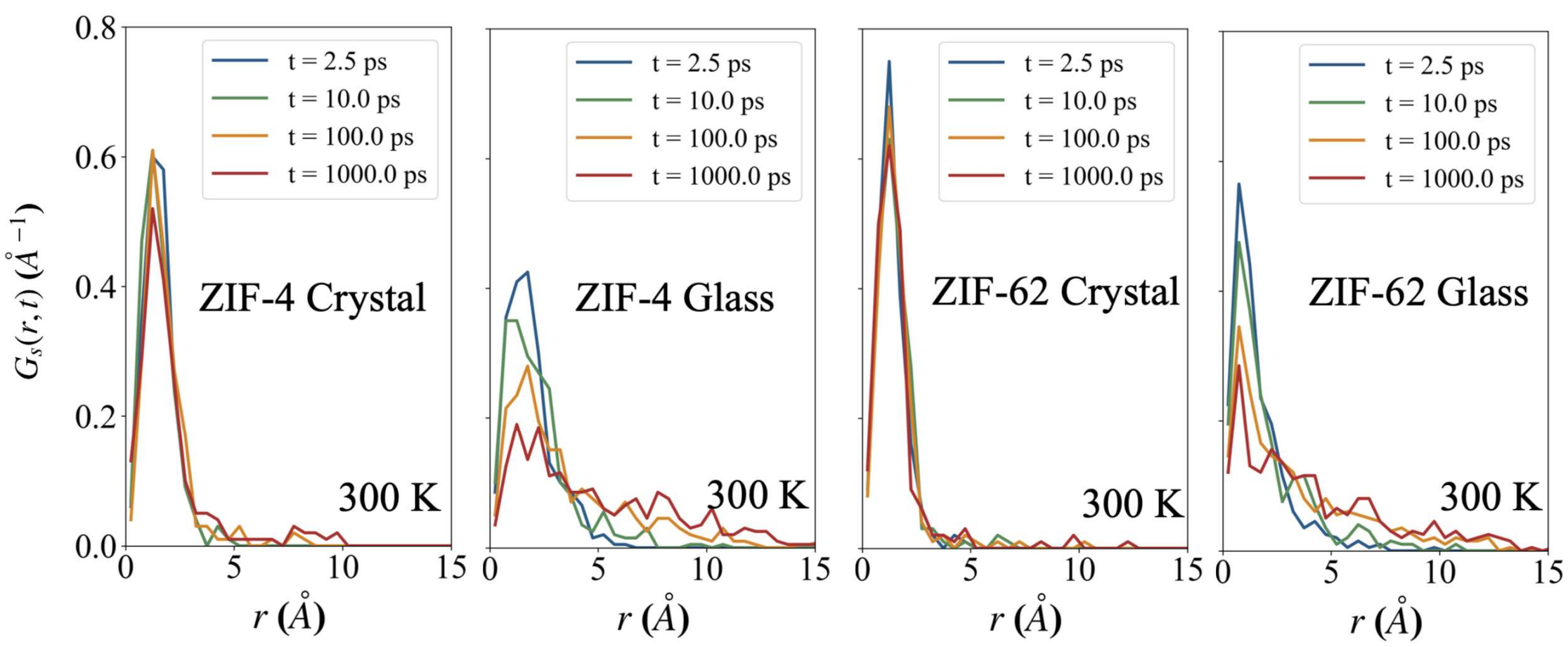

**Figure S4.** Evolution of $G_s(r,t)$ for the crystalline and glassy ZIF-4 and ZIF-62 samples at selected times of 2.5, 10.0, 100, and 1000 ps.

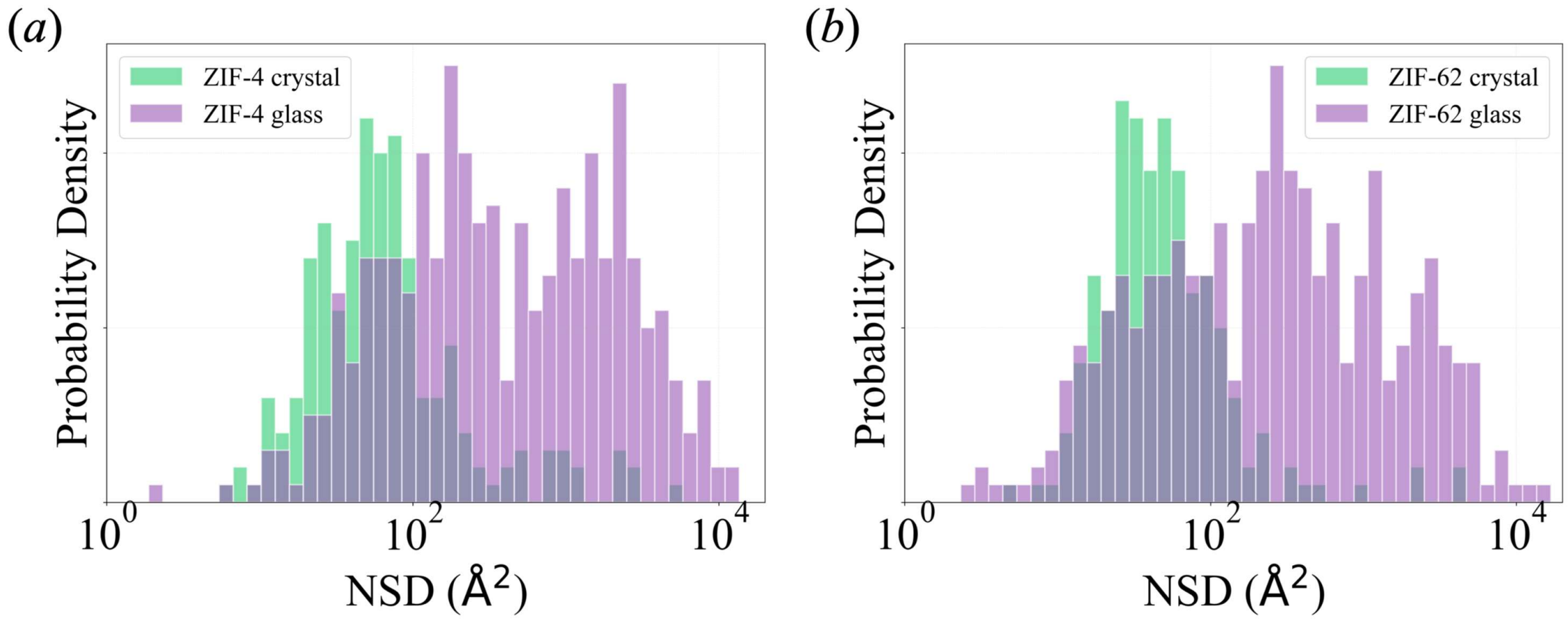

**Figure S5.** Illustration of probability density of non-affine square displacement of crystal/glassy (*a*) ZIF-4 and (*b*) ZIF-62 at a time scale of 1 *ns*.

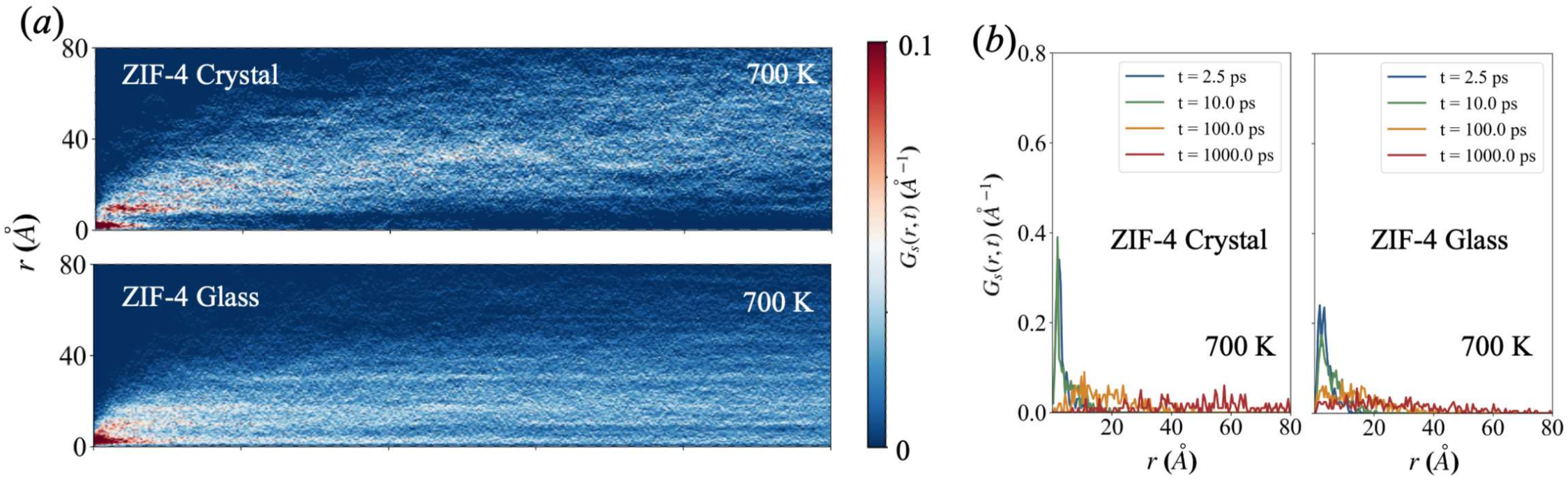

**Figure S6.** (*a*) Self-part van Hove correlation function $G_s(r,t)$ for both crystalline and glassy ZIF-4/ZIF-62 at a temperature of 700 K. (*b*) Evolution of $G_s(r,t)$ for the same systems at selected times of 2.5, 10.0, 100, and 1000 ps.

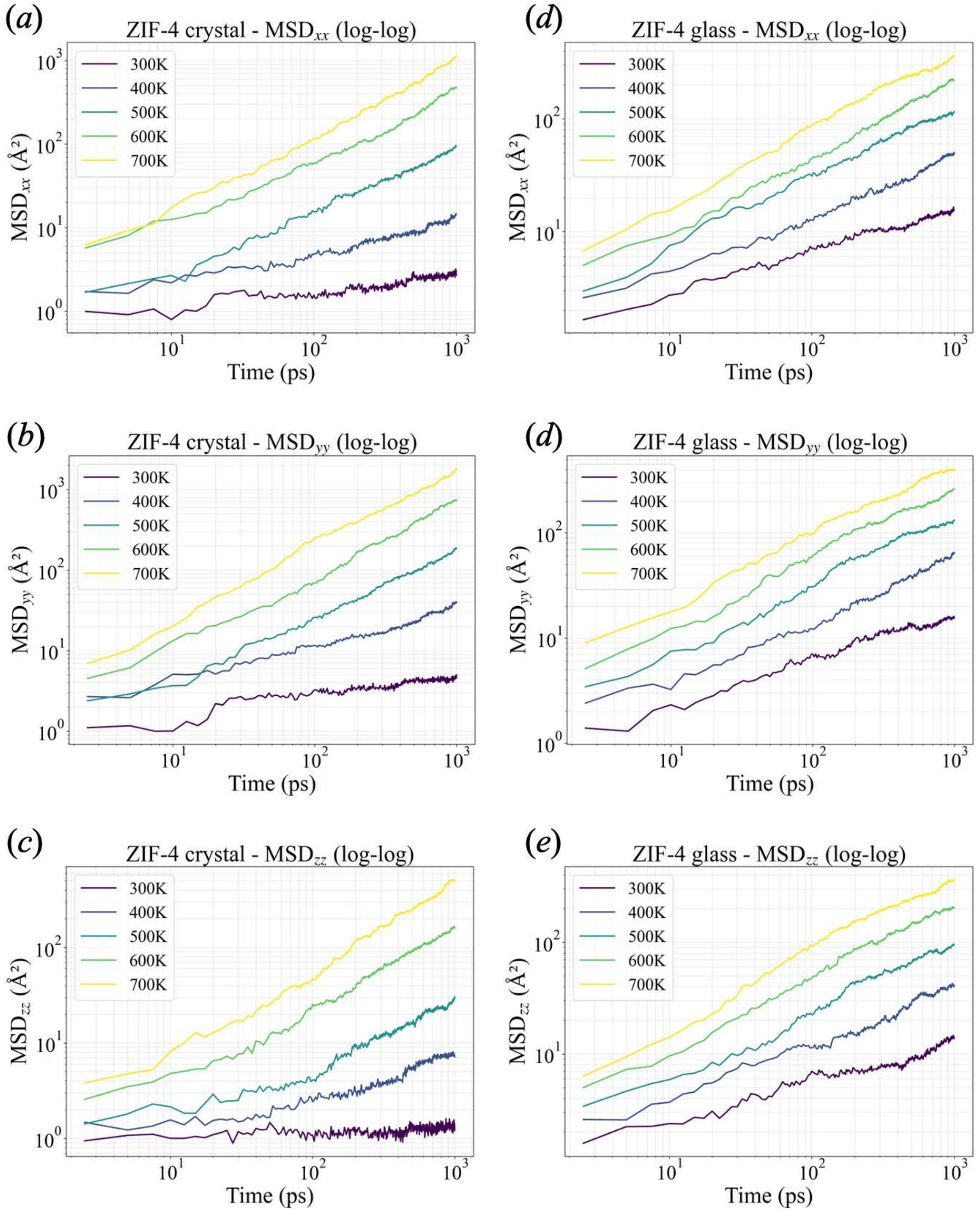

**Figure S7.** Directional mean squared displacements $MSD_{xx}$, $MSD_{yy}$ and $MSD_{zz}$ in ZIF-4 crystal (*a*-*c*) and ZIF-4 glass (*d*-*e*) at different temperatures.

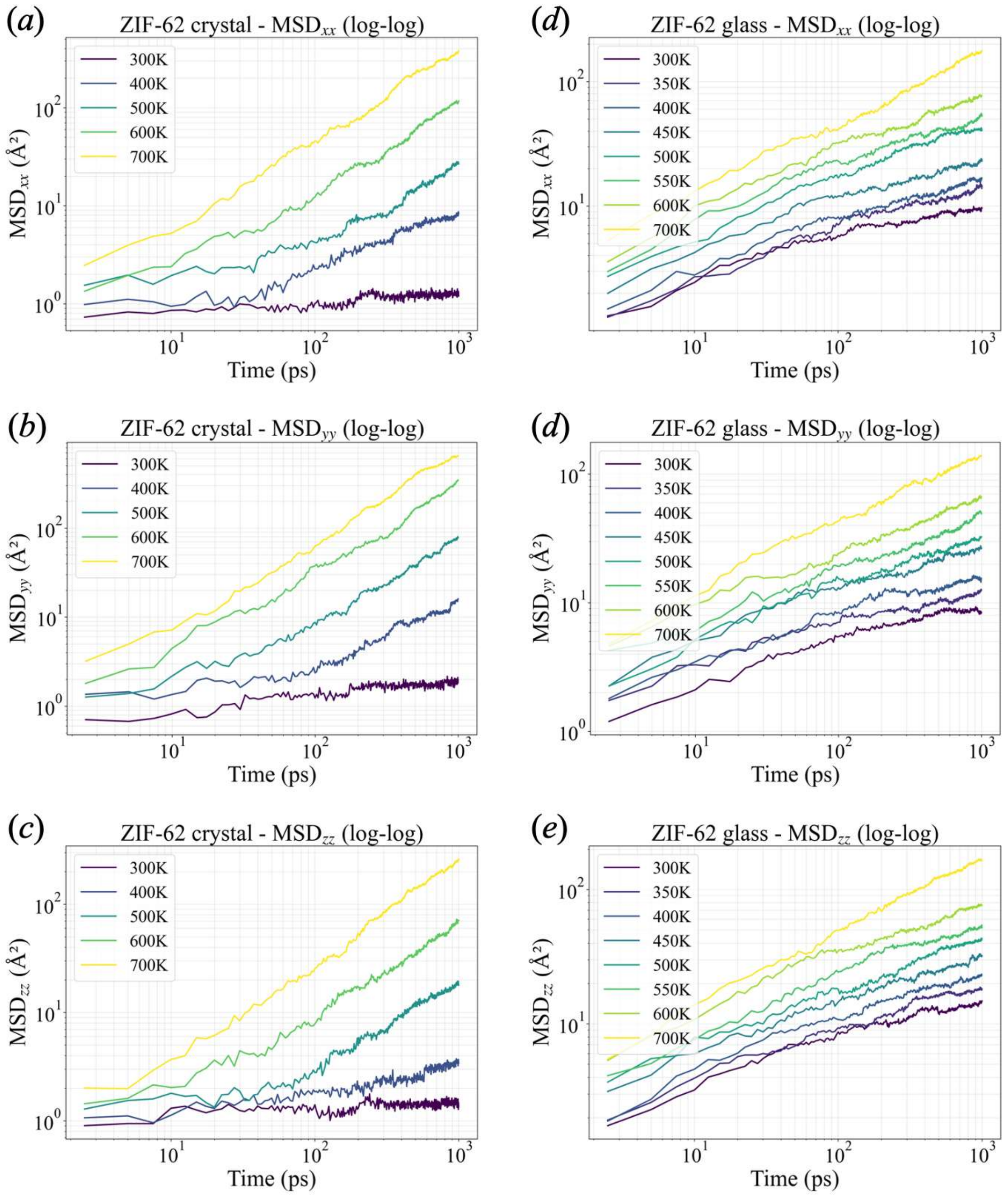

**Figure S8.** Directional mean squared displacements $MSD_{xx}$, $MSD_{yy}$ and $MSD_{zz}$ in ZIF-62 crystal (*a*-*c*) and ZIF-62 glass (*d*-*e*) at different temperatures.

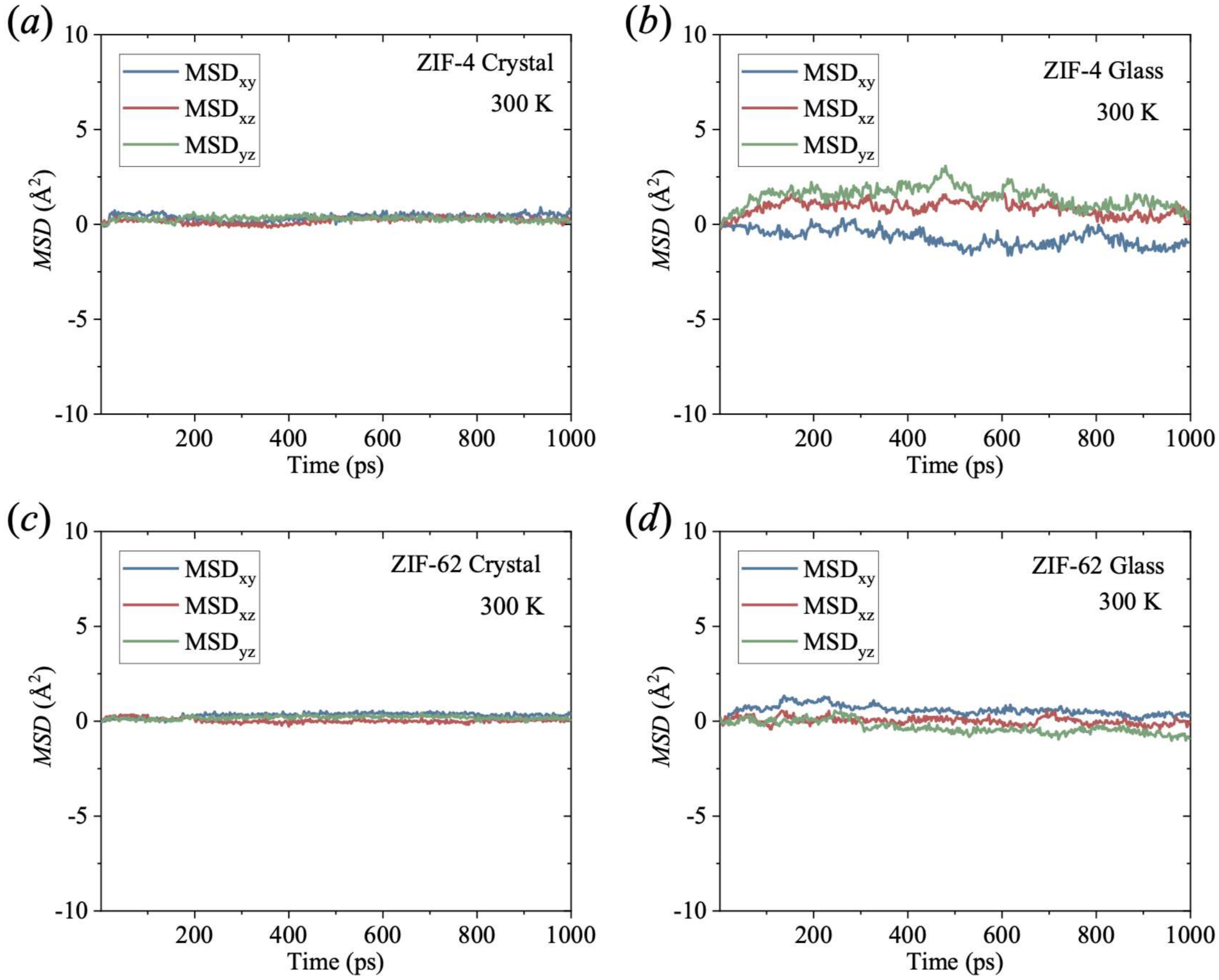

**Figure S9.** Cross terms of mean-squared displacement (MSD), i.e., $MSD_{xy}$, $MSD_{xz}$ and $MSD_{yz}$, in (*a*) ZIF-4 crystal, (*b*) ZIF-4 glass, (*c*) ZIF-62 crystal, and (*d*) ZIF-62 glass at room temperature.

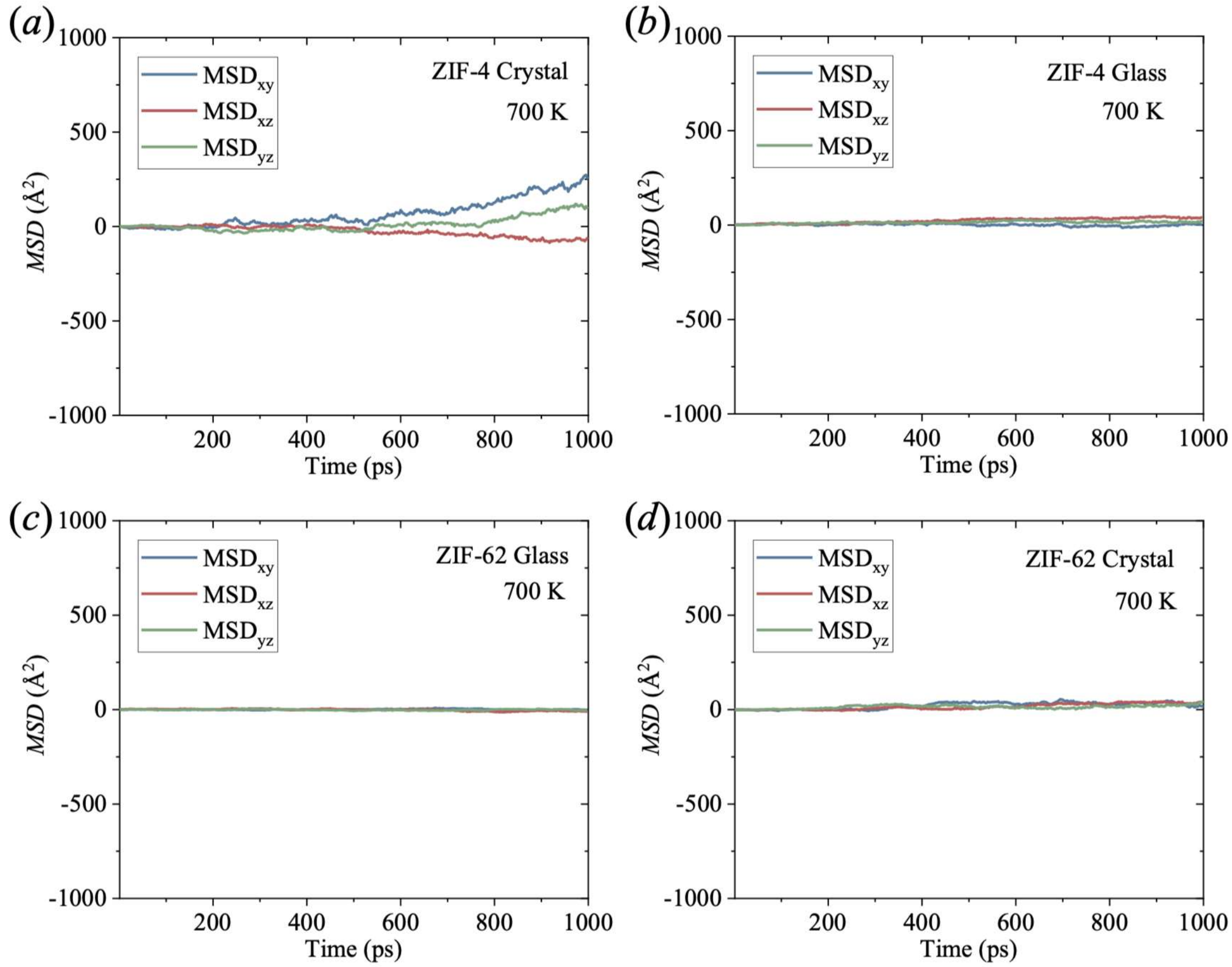

**Figure S10.** Cross terms of mean-squared displacement (MSD), i.e., $MSD_{xy}$, $MSD_{xy}$ and $MSD_{yz}$, in (*a*) ZIF-4 crystal, (*b*) ZIF-4 glass, (*c*) ZIF-62 crystal, and (*d*) ZIF-62 glass at a high temperature of 700 K.

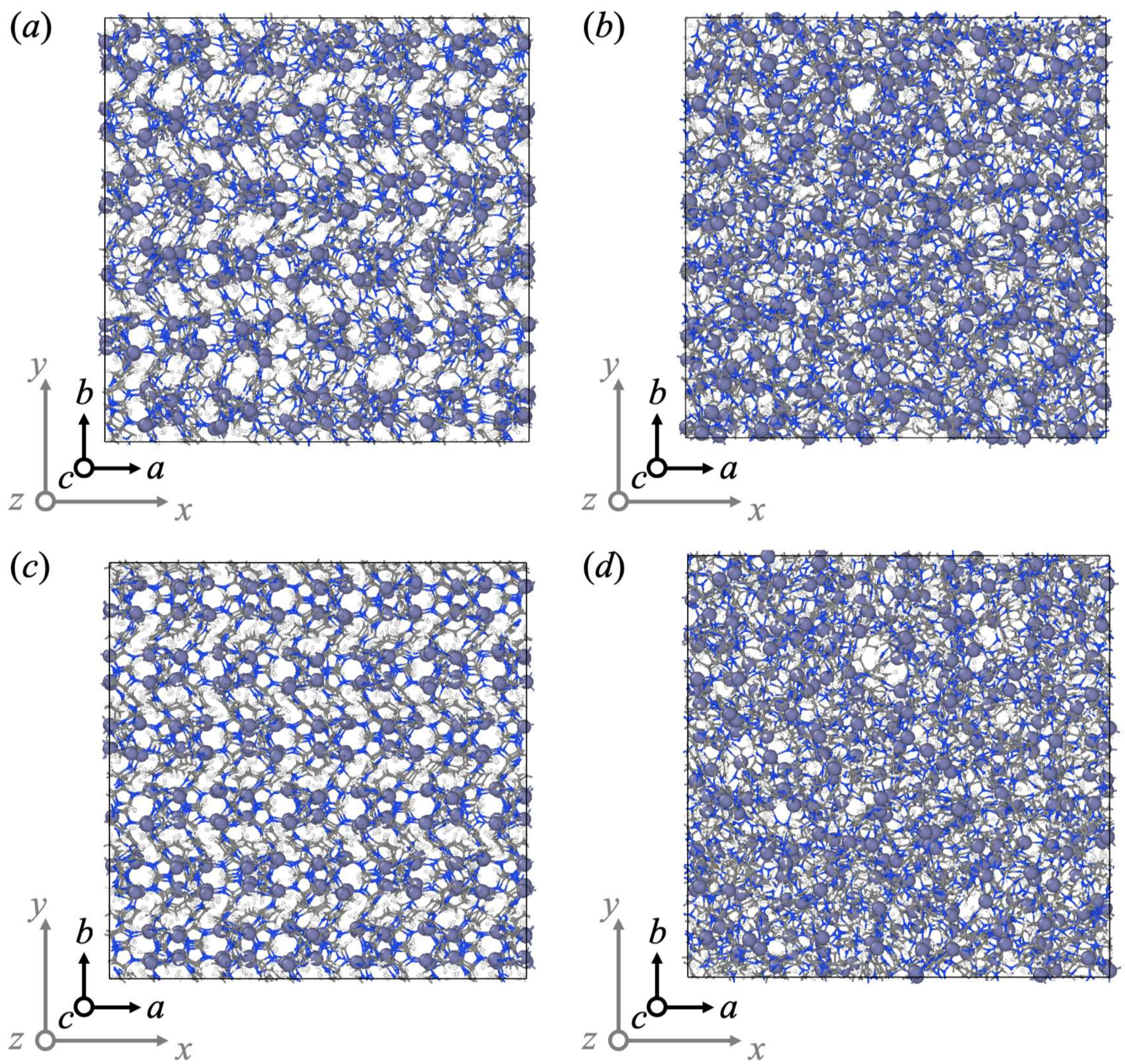

**Figure S11.** Atomic snapshots of the simulation samples and their corresponding lattice orientations in Cartesian coordinates: (*a*) ZIF-4 crystal, (*b*) ZIF-4 glass, (*c*) ZIF-62 crystal, and (*d*) ZIF-62 glass.

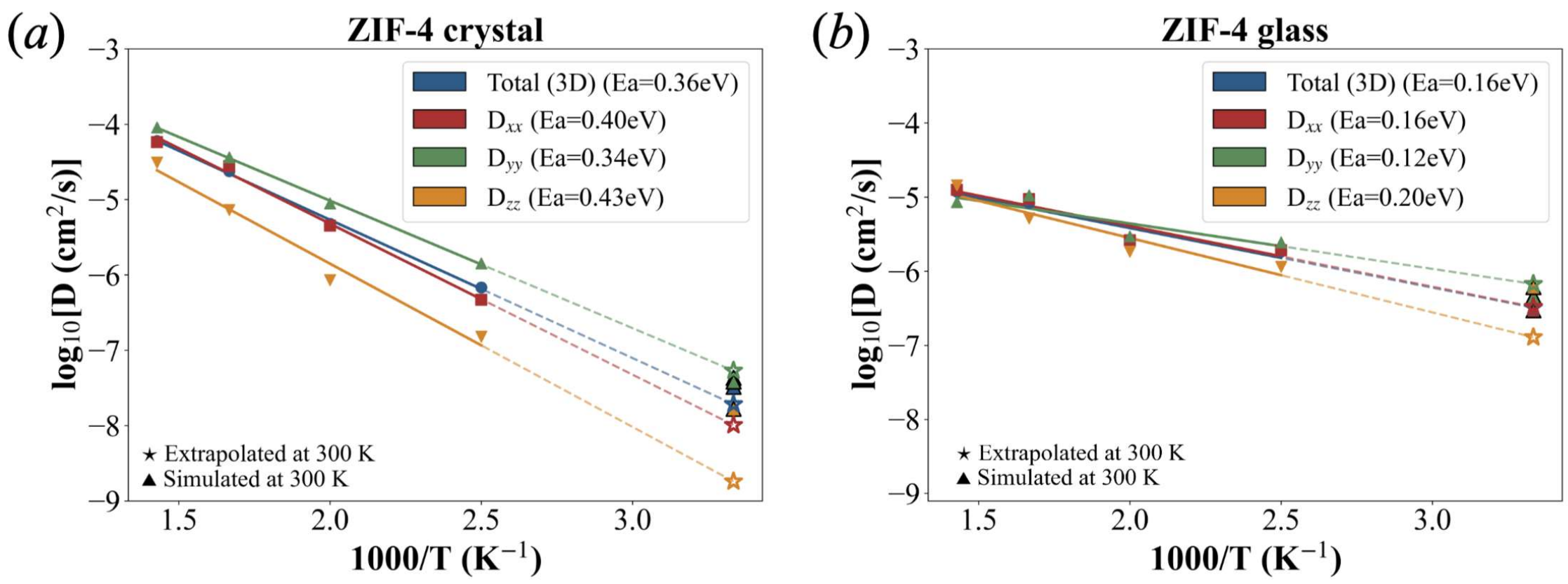

**Figure S12.** Temperature-dependent lithium-ion diffusion coefficients along the *x*, *y*, and *z* directions (denoted as $D_{xx}$, $D_{yy}$ and $D_{zz}$) for (*a*) crystalline ZIF-4 and (*b*) glassy ZIF-4.

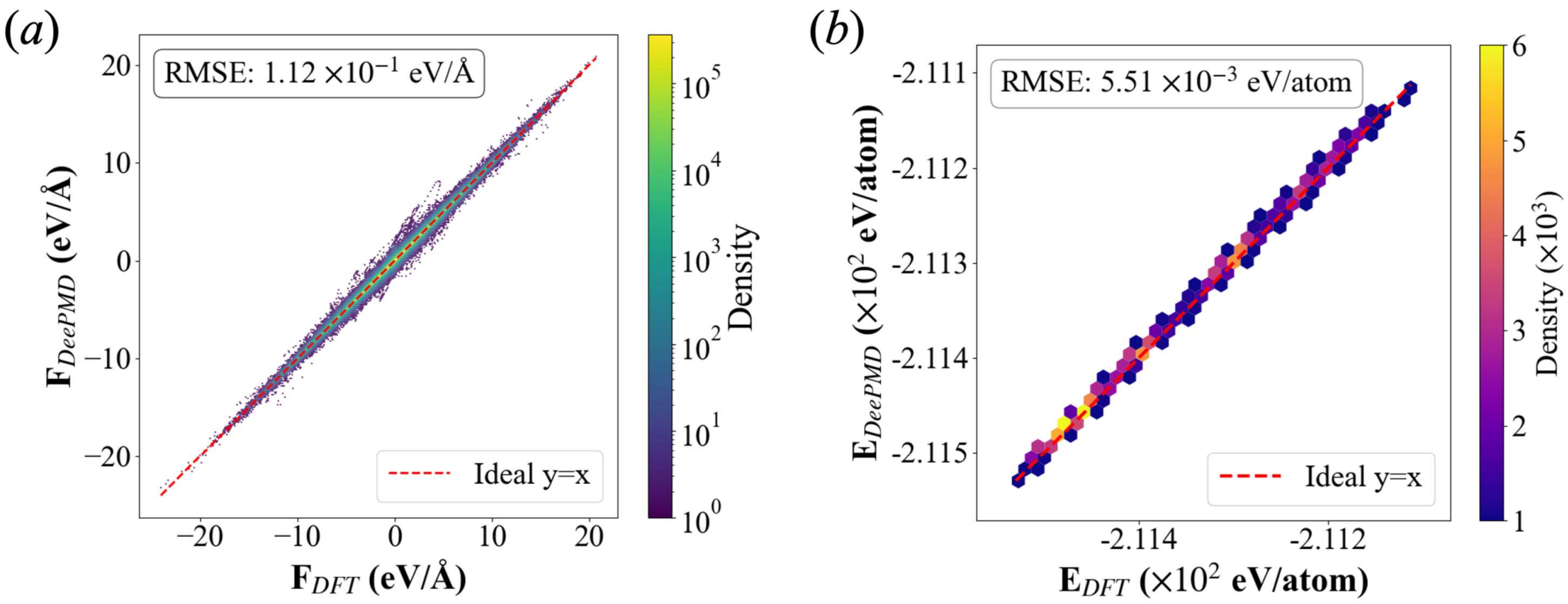

**Figure S13.** Parity plots comparing atomic forces (left panel) and energies (right panel) predicted by the DeePMD-kit MLIP with DFT reference values for a lithium ion inside ZIF-62 crystal at different temperatures: 300, 500, 1000, 2000, and 3000 K.